\documentclass[%
 reprint,
superscriptaddress,
 amsmath,amssymb,
 aps,
]{revtex4-1}

\usepackage{graphicx}
\usepackage{dcolumn}
\usepackage{bm}
\usepackage{bbm}
\usepackage{subfigure}
\usepackage{longtable,booktabs}
\usepackage{bbding}
\usepackage{verbatim}

\usepackage{color}
\definecolor{Blue}{rgb}{0.3,0.3,0.9}
\definecolor{Red}{rgb}{0.9,0.3,0.3}
\definecolor{Green}{rgb}{0.3,0.6,0.3}
\newcommand{\revision}[1]{{{#1}}}
\newcommand{\revisionB}[1]{{{#1}}}

\begin{document}


\title{Disorder effects in the two-dimensional Lieb lattice and its extensions}

\author{Xiaoyu Mao}
\email{Maoxiaoyu@smail.xtu.edu.cn}
\affiliation{School of Physics and Optoelectronics, Xiangtan University, Xiangtan 411105, China}
	
\author{Jie Liu}
\email{liujie@smail.xtu.edu.cn}
\affiliation{School of Physics and Optoelectronics, Xiangtan University, Xiangtan 411105, China}

\author{Jianxin Zhong}
\email{jxzhong@xtu.edu.cn}
\affiliation{School of Physics and Optoelectronics, Xiangtan University, Xiangtan 411105, China}

\author{Rudolf A. R\"omer}
\affiliation{School of Physics and Optoelectronics, Xiangtan University, Xiangtan 411105, China}
\email{r.roemer@warwick.ac.uk}
\affiliation{Department of Physics, University of Warwick, Coventry, CV4 7AL,
	United Kingdom}
\affiliation{
CY Advanced Studies and LPTM (UMR8089 of CNRS),
CY Cergy-Paris Universit\'{e},
F-95302 Cergy-Pontoise, France%
}

\date{\today}

\begin{abstract}

We study the localization properties of the two-dimensional Lieb lattice and its extensions in the presence of disorder using the transfer matrix method and finite-size scaling. We find that all states in the Lieb lattice and its extensions are localized for $W \geq 1$. Clear differences in the localization properties between disordered flat bands and disordered dispersive bands are identified. Our results complement previous experimental studies of clean photonic Lieb lattices and provide information about their stability with respect to disorder.
%
\end{abstract}

\maketitle


\section{\label{sec:intro}Introduction}


\revision{Quantum descriptions of materials in the solid state} can be carried out equally well in real space as in momentum space \cite{Ashcroft1976,Hook1991}. In perfect crystals, the momentum space description offers the advantage that all spatial information is already encapsulated in the unit cell, hence allowing for a dramatic reduction in the complexity of characterising such materials. 
The structure of the resulting energy bands, their number, curvature, linearity, filling and behaviour at the Fermi energy are all important descriptors, essential for the determination of the material properties of said systems. In recent years, e.g., interest has broadened this approach towards a description of the topological properties of the bands \cite{Kane2005,Bernevig2006} with concepts such as Fermi arcs \cite{Wan2011TopologicalIridates,Xu2015ObservationMetal} and Weyl points \cite{Soluyanov2015Type-IISemimetals} indicating highly-specialized band structure features. 

In a similar context, so called \emph{flat bands} have also received much attention \cite{Leykam2018Perspective:Flatbands}. The term refers to energy bands that are dispersionless in the whole of $k$-space \cite{Tasaki1998FromModel,Miyahara2007BCSLattice,Bergman2008BandModels,Wu2007FlatLattice}. This is equivalent to effectively having zero kinetic energy in these bands and hence to promoting the remaining terms in the Hamiltonian such as potential and interaction terms. Usually, such flat bands emerge only for specially "engineered" lattice structures such as quasi-1D lattices \cite{Leykam2017LocalizationStates,Shukla2018,Ramachandran2017}, diamond-type lattices \cite{Goda2006InverseFlatbands}, and Lieb lattices \cite{Qiu2016DesigningSurface,Julku2016GeometricBand,Chen2017Disorder-inducedLattices,Nita2013,Sun2018ExcitationLattice,Bhattacharya2019FlatLattice}.
Lieb lattices were originally used to enhance magnetic effects in model studies \cite{Lieb1989TwoModel,Mielke1993FerromagnetismModel,Tasaki1998FromModel} but have now been shown to be of experimental relevance in Wigner crystals \cite{Wu2007FlatLattice}, high-temperature superconductivity \cite{Miyahara2007BCSLattice,Julku2016GeometricBand}, photonic wave guide arrays \cite{Vicencio2015a,Mukherjee2015a,Guzman-Silva2014ExperimentalLattices,Diebel2016ConicalLattices,Leykam2018Perspective:Flatbands}, Bose-Einstein  condensates \cite{Baboux2016BosonicBand,Taie2015CoherentLattice}, ultra-cold atoms in optical lattices \cite{Shen2010SingleLattices} and electronic systems \cite{Slot2017ExperimentalLattice}.

The eigenstates in a flat band lattice are known to form so-called compact localized states \cite{Leykam2017LocalizationStates}, which due to the effectively zero group velocity, cannot participate in transport although they are in principle infinitely degenerate \cite{Goda2006InverseFlatbands}. Since the existence of flat bands relies on exact conditions, weak deviations from these conditions, such as potential disorder, could in principle fundamentally alter the nature of these localized states. Indeed, previous studies reveal a wide spread of tendencies depending on the nature of the perturbations \cite{Souza2009Flat-bandModel,Chalker2010AndersonBands,Nishino2007Flat-bandSystem,Goda2006InverseFlatbands,Flach2014DetanglingLattices,Leykam2017LocalizationStates} as well as the environment of
the bands \cite{Vidal2000InteractionPotential,Vidal2001DisorderCages,Gulacsi2004ExactInteractions,Gulacsi2010RoutePolymers}.
Our focus here is slightly different: instead of concentrating on the localization properties of the flat band states, we will investigate how the localization properties in the neighboring dispersive bands are changed by the presence of the flat bands. In particular, we are interested in whether the presence of the flat bands changes the localization properties and the finite-size scaling behaviour in disordered two-dimensional (2D) Lieb model and its extensions.

\section{\label{sec:model}Models and Method}

\subsection{\label{sec:lieb}The Lieb lattices $\mathcal{L}(n)$}

\begin{figure*}[tbh]
\centering   
$\mathcal{L}(1)$\includegraphics[width=0.43\columnwidth]{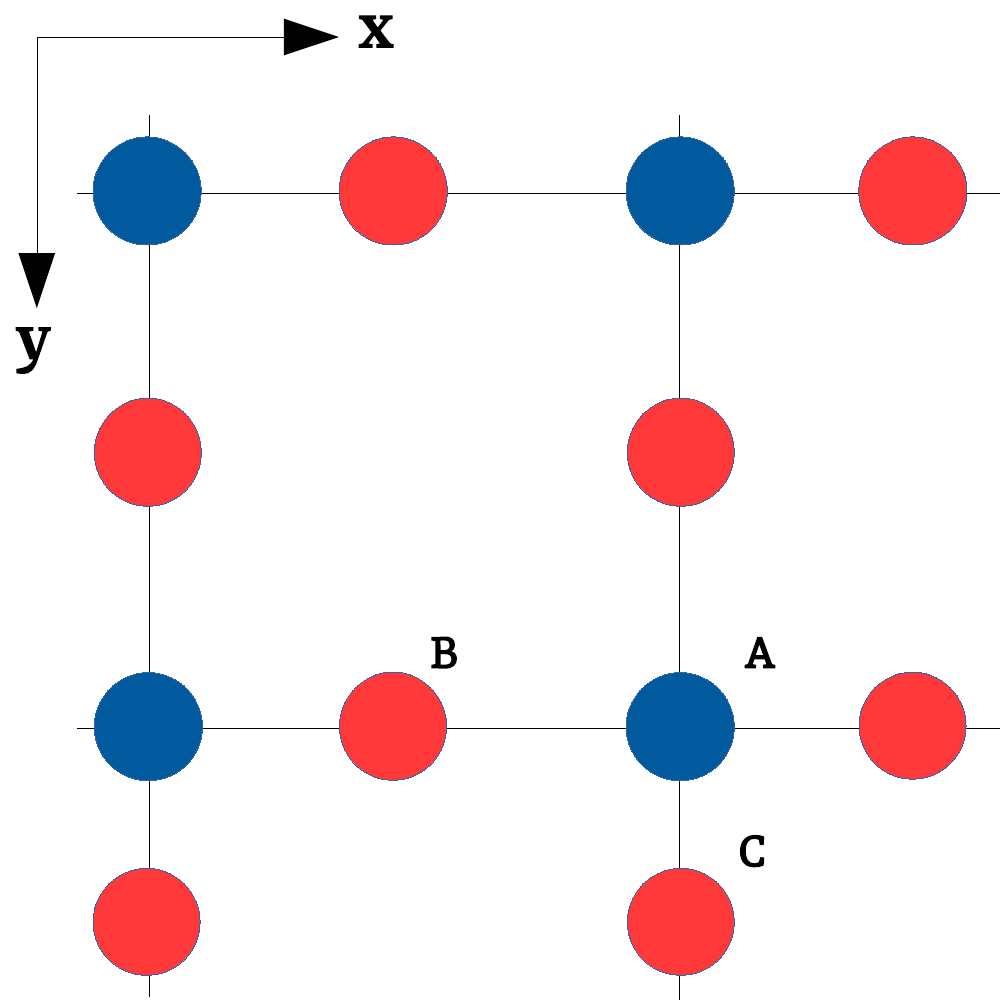} 
$\mathcal{L}(2)$\includegraphics[width=0.43\columnwidth]{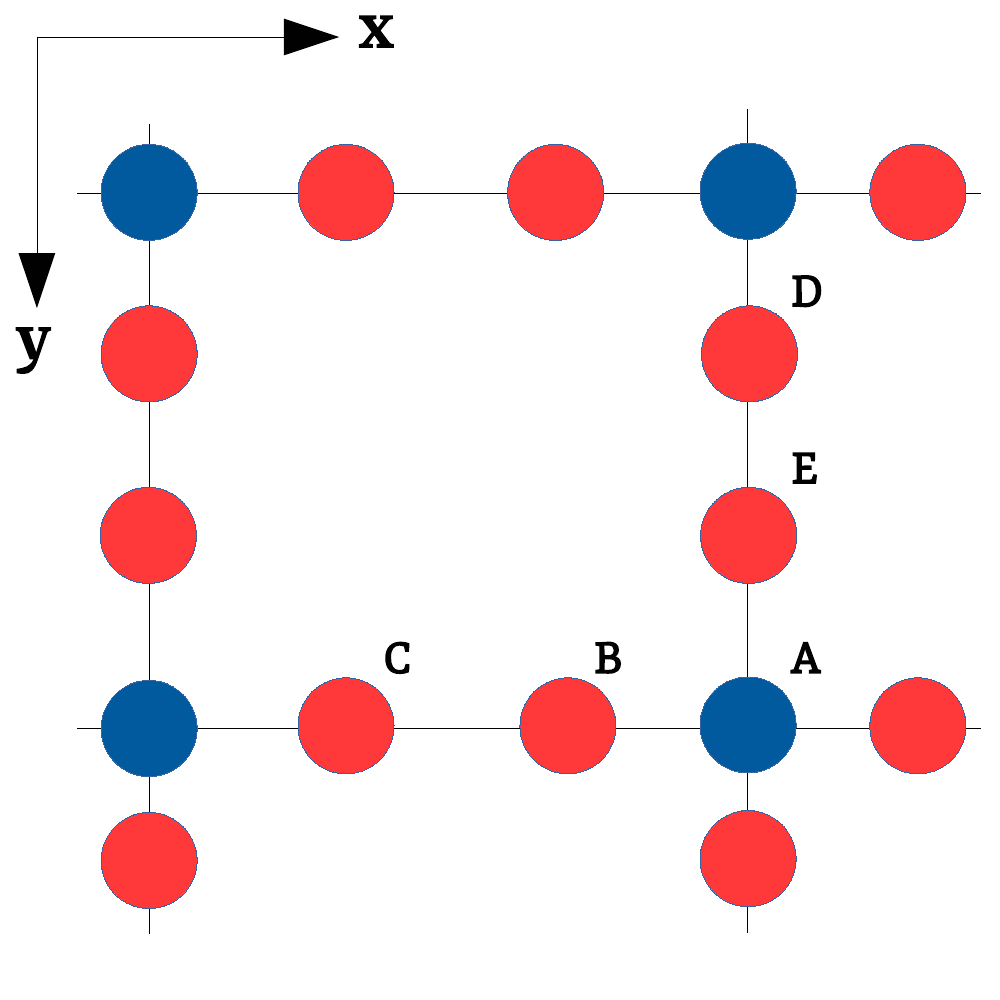} 
$\mathcal{L}(3)$\includegraphics[width=0.43\columnwidth]{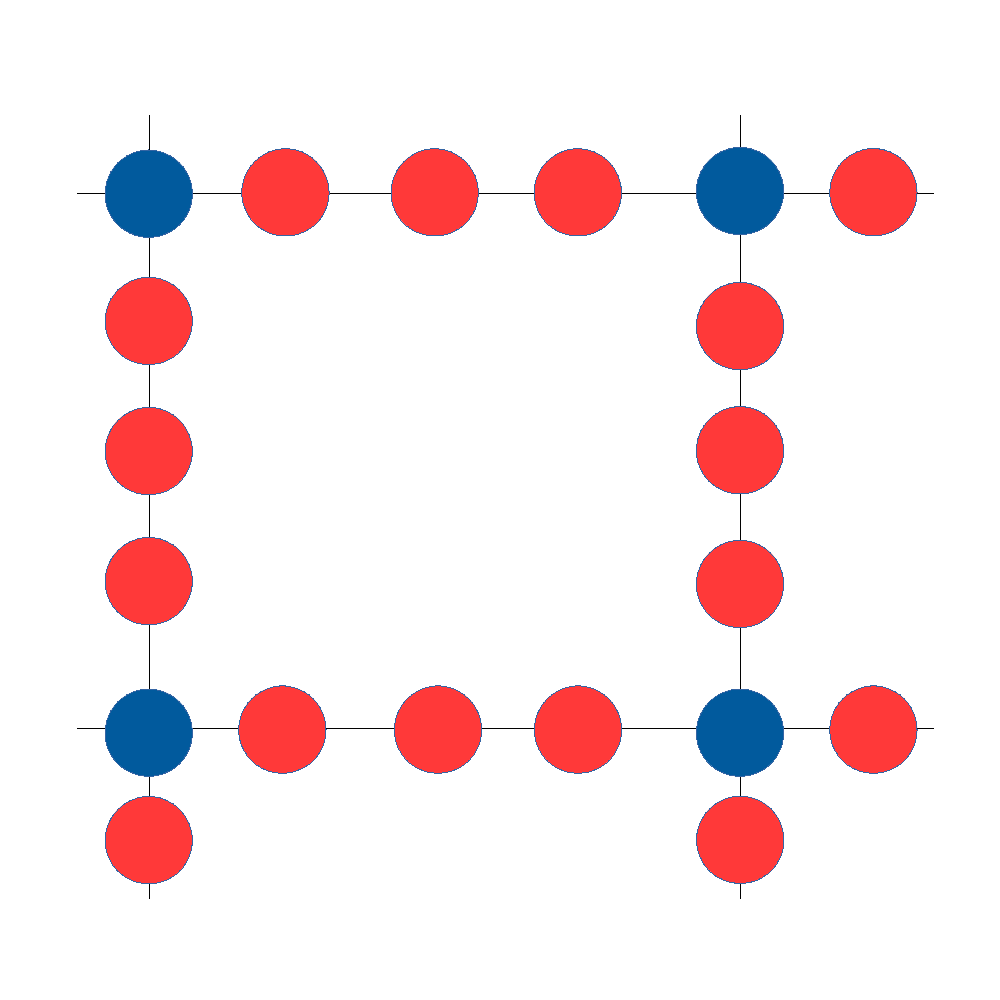}
$\mathcal{L}(4)$\includegraphics[width=0.43\columnwidth]{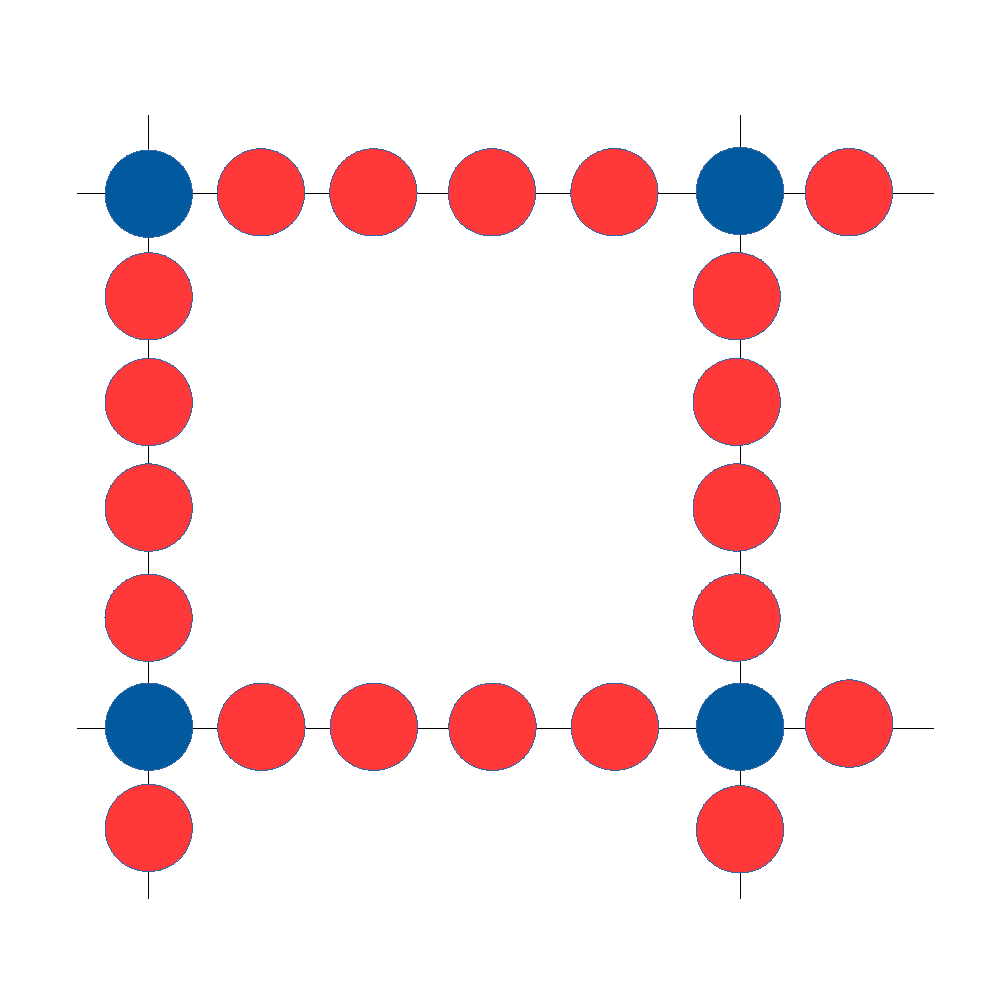} 
\caption{Schematic structure of lattices $\mathcal{L}(1)$, $\mathcal{L}(2)$, $\mathcal{L}(3)$ and $\mathcal{L}(4)$, respectively. Circles denote the lattice sites with blue/dark circles corresponding to the sites of a simple square lattice and the red/light circles denoting the added sites of the Lieb lattices. The thin black lines between the sites indicate hopping connections. \revision{The letters} A, B, \ldots, E give special labels used in the explanation of the TMM in the text. The $x$ and $y$ axes are given by arrows.
}
\label{fig:LiebLatticeStructure}                               
\end{figure*}

The original Lieb lattice \cite{Lieb1989TwoModel} is based on a square lattice with additional sites added along the horizontal and vertical directions as shown in Fig.\ \ref{fig:LiebLatticeStructure}. We indicate the original sites as A while B and C denote the added horizontal and vertical sites, respectively. In Ref.\ \onlinecite{Zhang2017NewBands}, extended Lieb lattices have been proposed with additional added sites as also shown in Fig.\ \ref{fig:LiebLatticeStructure}. We shall denote all these lattices symbolically as 
$\mathcal{L}(n)$ with $n$ marking the number of sites added in both horizontal and vertical directions, i.e.\ here $\mathcal{L}(1)$ to $\mathcal{L}(4)$.
%
Our numerical approach to study the localization properties of the Lieb lattices $\mathcal{L}(n)$ is based on (i) direct diagonalization of the Hamiltonian for small system sizes, and (ii) the transfer matrix method (TMM) \cite{MacKinnon1983a,Krameri1993}. TMM has been used extensively for various disorder problems \cite{Milde2000a,Cheraghchi2005Localization-delocalizationDisorder} and is the currently accepted method-of-choice for high-precision estimates of parameter-dependent localization lengths \cite{Slevin1999b}.

We start with the standard Anderson tight-binding Hamiltonian
\begin{equation}
    H = 
    \sum_{\vec{r}} 
    \varepsilon _{\vec{r}} \left| \vec{r} \right\rangle \left\langle \vec{r} \right|
    - 
    \sum_{\langle \vec{r} \ne \vec{r}'\rangle} 
    t_{\vec{r}, \vec{r}'} \left| \vec{r} \right\rangle \left\langle \vec{r}' \right|
    \label{eqn:Hamiltonian} ,
\end{equation}
where $| \vec{r} \rangle$ denotes the tight-binding state site $\vec{r}=(x,y)$, $x$ and $y$ indicate the exact position in the 2D plane and ${\langle \vec{r} \ne \vec{r}'\rangle}$ indicates the nearest-neighbor connections as defined in Fig.\ \ref{fig:LiebLatticeStructure}.
The onsite potential energy $\varepsilon_{\vec{r}}$ at site $\vec{r}$ is randomly and uniformly distributed in $[-W/2,W/2]$ such that $W$ indicates the strength of the disorder. 
The $t_{\vec{r}, \vec{r}'}$ are the hopping integrals for a particle going from site $\vec{r}$ to site $\vec{r}'$, and represent the kinetic energy part of the Hamiltonian. 
We only set nearest neighbours hopping in our model, so that $t_{\vec{r}, \vec{r}'}= t\equiv 1$ for nearest sites sets the energy scale and $t_{\vec{r}, \vec{r}'} = 0$ otherwise. \revision{We have chosen hard-wall boundary conditions, but also checked that results are similar for periodic boundaries.}

\subsection{\label{sec:tmm-l1}Transfer-matrix method for $\mathcal{L}(1)$}

For the $\mathcal{L}(1)$ lattice, in order to compute the localization length $\lambda$ of the wave function from the TMM corresponding to the Schr\"{o}dinger equation $\mathbf{H}\Psi = E\Psi$, we consider a quasi-1D strip consisting of $M$ chains of A and B \revision{sites with additional $M$ sites, labelled C,} in the vertical ($y$-) direction \revision{as shown in Fig.\ \ref{fig:LiebLatticeStructure}. We choose to }propagate the TMM in the horizontal ($x$-) direction until convergence \revisionB{\footnote{With typical convergence error of $0.1\%$ for $\mathcal{L}(1)$ this implies lengths of TMM bars of width $M=20$ up to $L=6\times 10^7$ layers at $W=1$, with a target error of $0.2\%$ for $\mathcal{L}(4)$ we find $L>10^8$ at $E=0$. The overall system sizes studied here are given by $(2n+1)M L$ for $\mathcal{L}(n)$.}}. This implies a TMM setup with \revision{two transfer matrices, $\mathbf{T}_{A \to B}$ and $\mathbf{T}_{B \to A}$. From sites A to B in $x$-direction, we find}
\begin{widetext}
\begin{equation}
\begin{aligned}
 \left( {\begin{array}{*{20}{c}}
{\psi _{x + 1}^B}\\
\\
{\psi _{x}^A}
\end{array}} \right) &=\mathbf{T}_{A \to B}\left( {\begin{array}{*{20}{c}}
{\psi _{x}^A}\\
\\
{\psi _{x - 1}^B}
\end{array}} \right)
=\left( {\begin{array}{{cr}}
{
\left(
\frac{{{\varepsilon _{x,y}} - E}}{t} - \frac{t}{{{\varepsilon _{x,y - 1}} - E}} - \frac{t}{{{\varepsilon _{x,y + 1}} - E}}
\right) \mathbf{1}_M
- \frac{ \mathbf{t}_{y} }{{{\varepsilon _{x,y - 1}} - E}} 
- \frac{ \mathbf{t}_{y}^{\dag}}{{{\varepsilon _{x,y + 1}} - E}}} & {\mbox{ } \quad -\mathbf{1}_M} \\
\\
\mathbf{1}_M & \mathbf{0}_M
\end{array}} \right)
\left( {\begin{array}{*{20}{c}}
{\psi _{x}^A}\\
\\
{\psi _{x - 1}^B}
\end{array}} \right)      
\end{aligned}
\label{eq:tmm-21}
\end{equation}
\end{widetext} 
with $M \times M$ matrix
\begin{equation}
  \mathbf{t}_{y}=t 
  \left(
  \begin{array}{ccccccccc}
 0 & 1 & 0  &\cdots&0 &0\\
 0 & 0 & 1  &\cdots & 0&0\\
  \vdots &  &  \ddots & & & \vdots\\
 0 & 0 & 0 &\cdots & 1 & 0\\
 0 & 0 & 0  &\cdots &0 &  1\\
 (1) & 0 & 0 &\cdots &0 & 0\\
  \end{array}
  \right) 
\end{equation}
denoting the hopping connections to neighboring A sites in $y$ direction and $(1)$ indicating a possible periodic boundary term in $y$-direction. Note that in this setup, the C sites have been effectively renormalized away into effective onsite energies and changed vertical hopping terms such that the TM has the standard $2M \times 2M$ size with $\mathbf{1}_M$ and $\mathbf{0}_M$ denoting $M\times M$ identity and zero matrices.
\revision{The $M \times M$ wave function amplitudes in the $x$th line are marked as $\Psi^{A,B}_{x}$, either A or B, with $y=1, \ldots, M$, labelling the position of the \emph{renormalized} cubic sites \cite{psiAD}.
In this notation the term $\frac{\varepsilon_{x,y}-E}{t} \mathbf{1}_{M} \equiv \mathrm{diag}\left(\frac{\varepsilon_{x,1}-E}{t},\frac{\varepsilon_{x,2}-E}{t}, \ldots, \frac{\varepsilon_{x,M}-E}{t}\right)$ while $y-1$ and $y+1$ indicate the \emph{unrenormalized} sites neighboring $y$ for the hopping terms with $\textbf{t}_{y}$ and $\textbf{t}_{y}^{\dag}$ in Eq.\ \eqref{eq:tmm-21}.}
From sites B to A, a more standard TMM form emerges as
\begin{eqnarray}
\lefteqn{
\left( {\begin{array}{*{20}{c}}
{\Psi _{x + 1}^A}\\
\\
{\Psi _{x}^B}
\end{array}} \right) = 
\mathbf{T}_{B\to A}\left( {\begin{array}{*{20}{c}}
{\Psi _{x}^B}\\
\\
{\Psi _{x - 1}^A}
\end{array}} \right)} \nonumber \\
\mbox{ }
&= & \left( {\begin{array}{*{20}{c}}
{{\left (\frac{{\varepsilon _{x,y}}- E}{t} \right) \mathbf{1}_M}} & { -\mathbf{1}_M}\\
\\
\mathbf{1}_M & \mathbf{0}_M
\end{array}} \right)\left( {\begin{array}{*{20}{c}}
{\Psi _{x}^B}\\
\\
{\Psi _{x - 1}^A}
\end{array}} \right)   .
\end{eqnarray}
%
The transfer matrices $\mathbf{T}_{B\to A}$ and $\mathbf{T}_{A\to B}$ transfer the wave vector amplitudes $\Psi$ between atoms A and B. Along the transfer ($x$-) axis, we multiply these matrices, reorthogonalize the $M$ $\Psi$ states, i.e.\ the columns of the TMM, at least after every 10th multiplication, compute Lyapunov exponents $\gamma_i$, $i=1, \ldots, M$ and study their accumulated changes in variance until we reach a desired accuracy (error of typically $0.1\%$) for the smallest $\gamma_\text{min}$ \cite{Krameri1993,Oseledets1968ASystems,Ishii1973LocalizationSystem,Beenakker1997Random-matrixTransport}. 
The localization length is then found as $\lambda(M,W,E)=1/\gamma_\text{min}$ for given energy $E$ and disorder strength $W$.

\subsection{\label{sec:tmm-l2}TMM for $\mathcal{L}(2)$ and beyond}

For the extended Lieb lattice $\mathcal{L}(2)$, we can derive the corresponding TMM recursion formula in a similar fashion. Starting again from the first sites in $x$-direction as given in Fig.\ \ref{fig:LiebLatticeStructure} for $\mathcal{L}(2)$, we find for the transfer from sites $A \to C $,
\begin{widetext}
\begin{equation}
\begin{aligned}
     \left(
     \begin{array}{c}
        \Psi_{x+1}^{C} \\
        \\
        \Psi_{x}^{A} 
     \end{array}
     \right)
     & = T_{A \to C}
      \left(
     \begin{array}{c}
        \Psi_{x}^{A} \\
        \\
        \Psi_{x-1}^{B} 
     \end{array}
     \right) \\
     & = \left(
     \begin{array}{crc}
        \mathcal{E}\mathbf{1}_M - 
            \frac{t \mathbf{t}_{y}}{\left(\varepsilon_{x,y-2}-E\right)\left(\varepsilon_{x,y-1}-E\right)-t^2}  -
            \frac{t \mathbf{t}^{\dag}_{y}}{\left(\varepsilon_{x,y+2}-E\right)\left(\varepsilon_{x,y+1}-E\right)-t^2}  &       \quad -\mathbf{1}_M\\
        \\
        \mathbf{1}_M &    \mathbf{0}_M&      
     \end{array}
     \right)
     \times
     \left(
     \begin{array}{c}
        \Psi_{x}^{A} \\
        \\
        \Psi_{x-1}^{B} 
     \end{array}
     \right),
\end{aligned} 
\end{equation}
where 
\begin{equation}
\begin{aligned}
  \mathcal{E} &= 
  \frac{\varepsilon_{x,y}-E}{t}
            -\frac{t\left(\varepsilon_{x,y-2}-E\right)}{\left(\varepsilon_{x,y-2}-E\right)\left(\varepsilon_{x,y-1}-E\right)-t^2}
            -\frac{t\left(\varepsilon_{x,y+2}-E\right)}{\left(\varepsilon_{x,y+2}-E\right)\left(\varepsilon_{x,y+1}-E\right)-t^2} 
            .
\end{aligned}
\end{equation}
\end{widetext}
From sites $C \to B$, we have
\begin{subequations}
\begin{eqnarray}
\lefteqn{
\left(
     \begin{array}{c}
        \Psi_{x+1}^{B} \\
        \\
        \Psi_{x}^{C} 
     \end{array}
     \right)
      =\mathbf{T}_{C\to B}
     \left(
     \begin{array}{c}
        \Psi_{x}^{C} \\
        \\
        \Psi_{x-1}^{A} 
     \end{array}
     \right) } \nonumber  \\
      &= &\left(
     \begin{array}{crc}
     (\frac{{\varepsilon _{x,y}}- E}{t} )\mathbf{1}_M &   & -\mathbf{1}_M \\
      \\
        \mathbf{1}_M                &   & \mathbf{0}_M
     \end{array}
     \right)
     \left(
     \begin{array}{c}
        \Psi_{x}^{C} \\
        \\
        \Psi_{x-1}^{A} 
     \end{array}
     \right)
 \end{eqnarray}
and for $B \to A$
\begin{eqnarray}
\lefteqn{
    \left(
     \begin{array}{c}
        \Psi_{x+1}^{A} \\
        \\
        \Psi_{x}^{B} 
     \end{array}
     \right)
     = \mathbf{T}_{B\to A}
     \left(
     \begin{array}{c}
        \Psi_{x}^{B} \\
        \\
        \Psi_{x-1}^{C} 
     \end{array}
     \right)} \nonumber \\
     &= &\left(
     \begin{array}{crc}
        (\frac{{\varepsilon _{x,y}}- E}{t} )\mathbf{1}_M &&-\mathbf{1}_M\\
        \\
        \mathbf{1}_M                && \mathbf{0}_M
     \end{array}
     \right)
     \left(
     \begin{array}{c}
        \Psi_{x}^{B} \\
        \\
        \Psi_{x-1}^{C} 
     \end{array}
     \right) .
\end{eqnarray}
\end{subequations}
As before, the hopping in vertical ($y$-) direction has been included in renormalization of the onsite energies of the A sites as well as renormalized hopping between A sites, albeit more complicated in structure than for $\mathcal{L}(1)$. Still, following this procedure, it is not difficult --- albeit time consuming --- to construct the transfer matrices for the remaining $\mathcal{L}(3)$ and $\mathcal{L}(4)$ lattices and we will not include those lengthy terms here.

\subsection{\label{sec:fss}Finite-size scaling in 2D}

According to renormalization group ideas \cite{Wilson1974TheExpansion}, we can study the properties of large critical systems from smaller subsystems \cite{Wegner1976ElectronsEdge,Lee1979Real-spaceLocalization,Sarker1981ScalingApproach}. We define the dimensionless, reduced localization length $\Lambda_{M}(E,W)=\lambda(E,W)/M$. In order to get the infinite-size localization length $\xi$ \cite{Leadbeater1999} from the $\Lambda_{M}(E,W)$, we should apply the one-parameter-scaling hypothesis, i.e.\ 
\begin{equation}
 \Lambda_{M}(E,W)=f(\xi(E,W)/M) .
\end{equation}
As a numerical procedure, this means that the finite-size $\Lambda_M$ values can be scaled by a factor $\xi$, such that all values collapse onto a single scaling function $f$. We fit the parameter $\xi$ by using a least squares procedure \cite{MacKinnon1983a} either as a function of $E$ for constant $W$, or, more standard, as a function of $W$ for constant $E$. We then determine the absolute scale of $\xi(W,E)$ by fitting the largest disorder with the form of a truly localised behaviour, i.e.\
\begin{equation}
    \Lambda_{M} = \xi/M + b(\xi/M)^2,
\end{equation}
as expected for large $W\gg t$.
We emphasize that numerically, this procedure does not use the Taylor expansion methods nor does it assume any power-law diverging terms as is normally standard in studies of the 3D Anderson transition and its variants \cite{Slevin1999b,Rodriguez2011MultifractalTransition}. Rather, as there is no \revision{evident metal-insulator transition in the 2D Anderson model \cite{Krameri1993}}, there is no universal functional form to fit towards \cite{MacKinnon1983a} and we hence have to revert back to the simpler methods of maximizing data overlap as done in most pre-1999 studies of Anderson localization \cite{Krameri1993}.

\begin{figure*}[tb]
    \centering
    $\mathcal{L}(1)$\includegraphics[width=0.43\columnwidth]{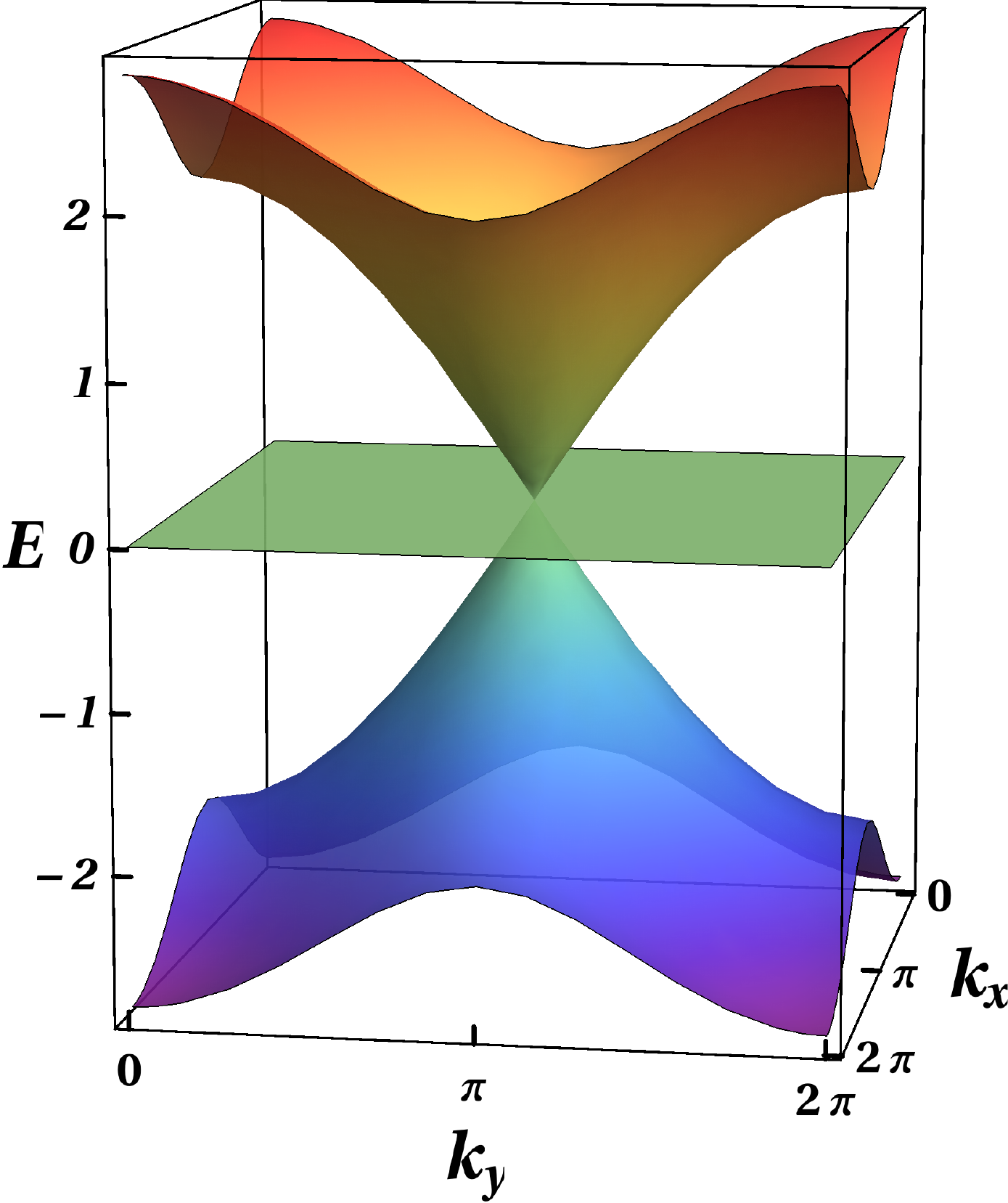}
    $\mathcal{L}(2)$\includegraphics[width=0.43\columnwidth]{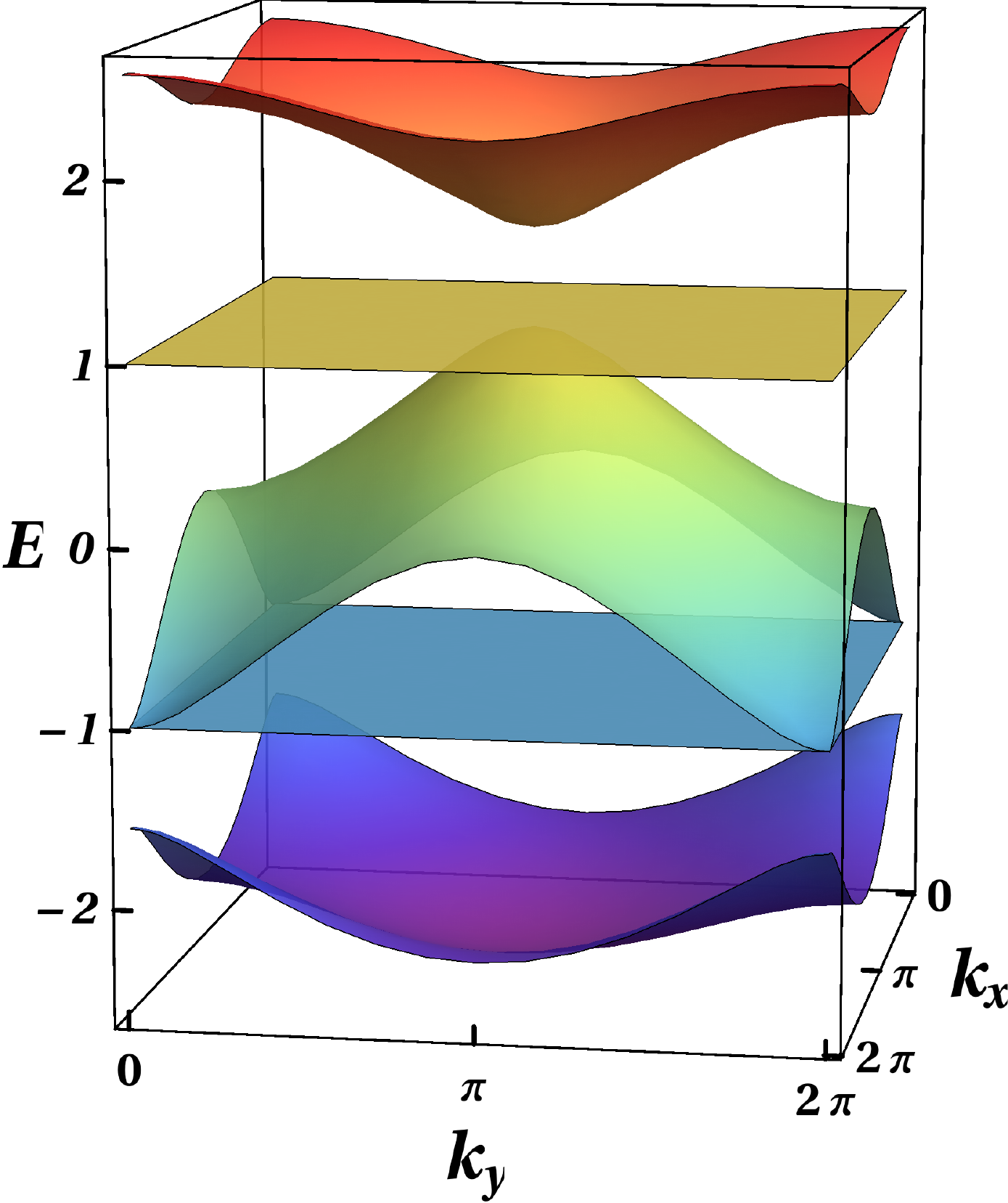}
    $\mathcal{L}(3)$\includegraphics[width=0.43\columnwidth]{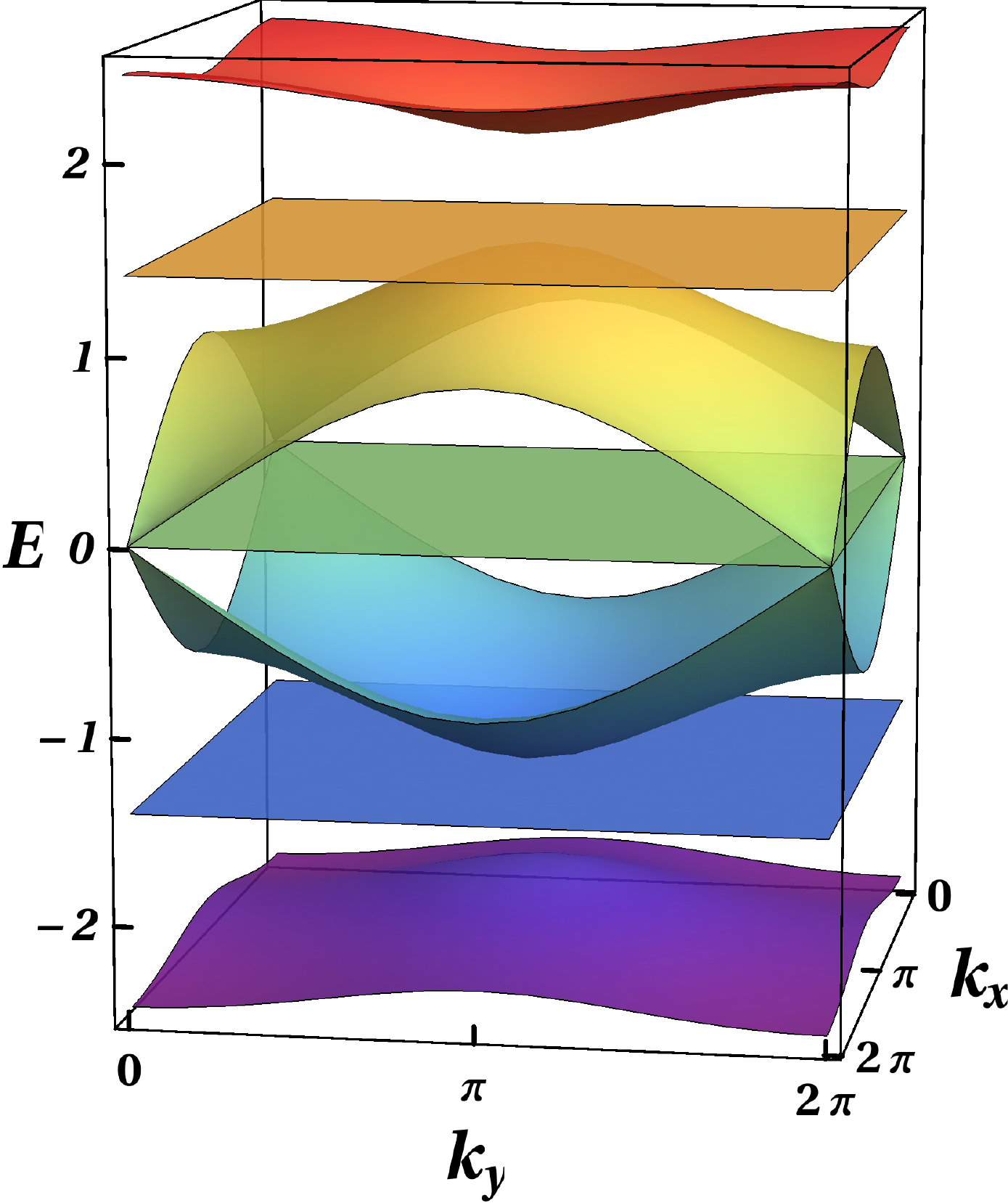}
    $\mathcal{L}(4)$\includegraphics[width=0.43\columnwidth]{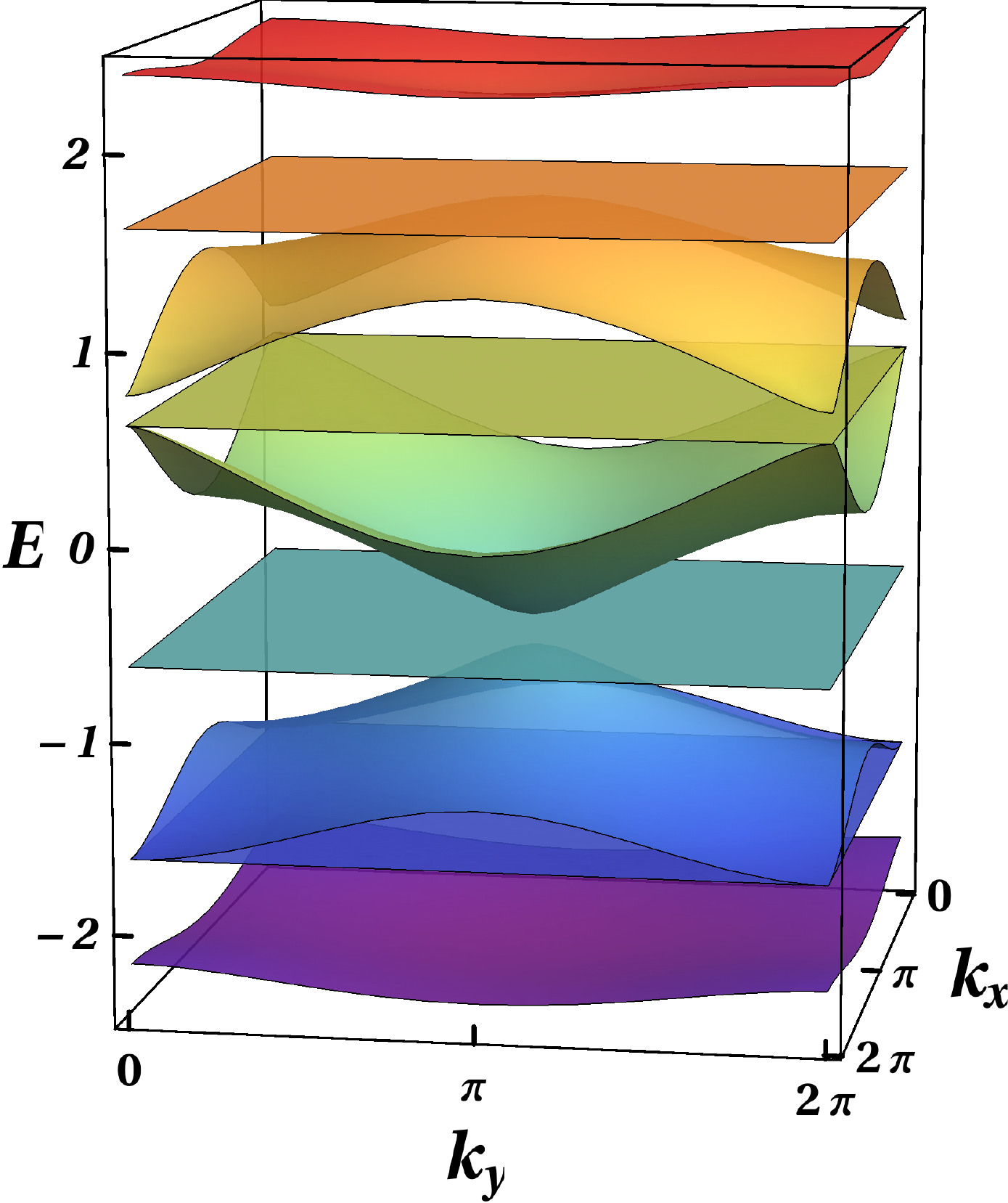}\\
    \includegraphics[width=0.49\columnwidth]{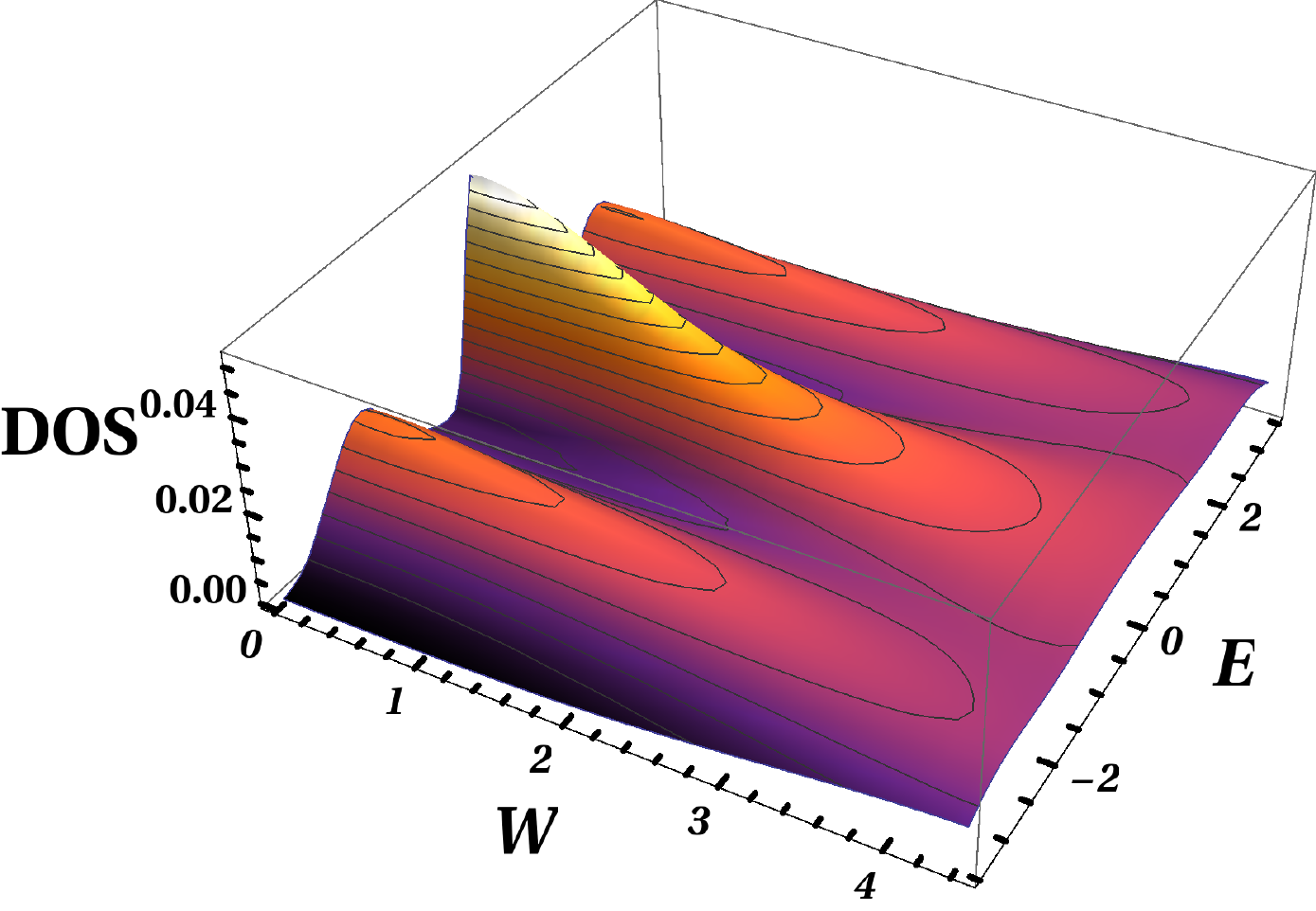}
    \includegraphics[width=0.49\columnwidth]{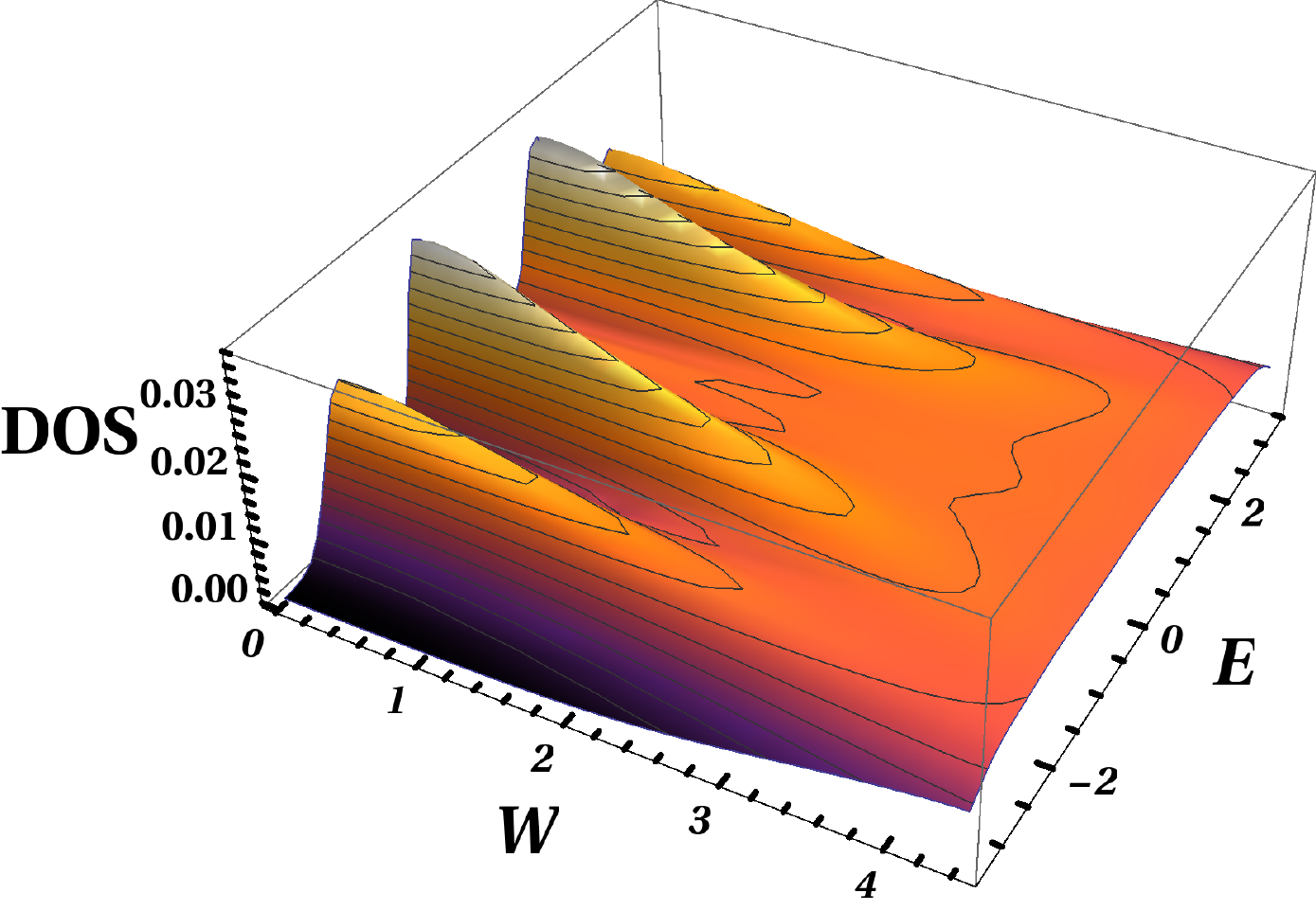}
    \includegraphics[width=0.49\columnwidth]{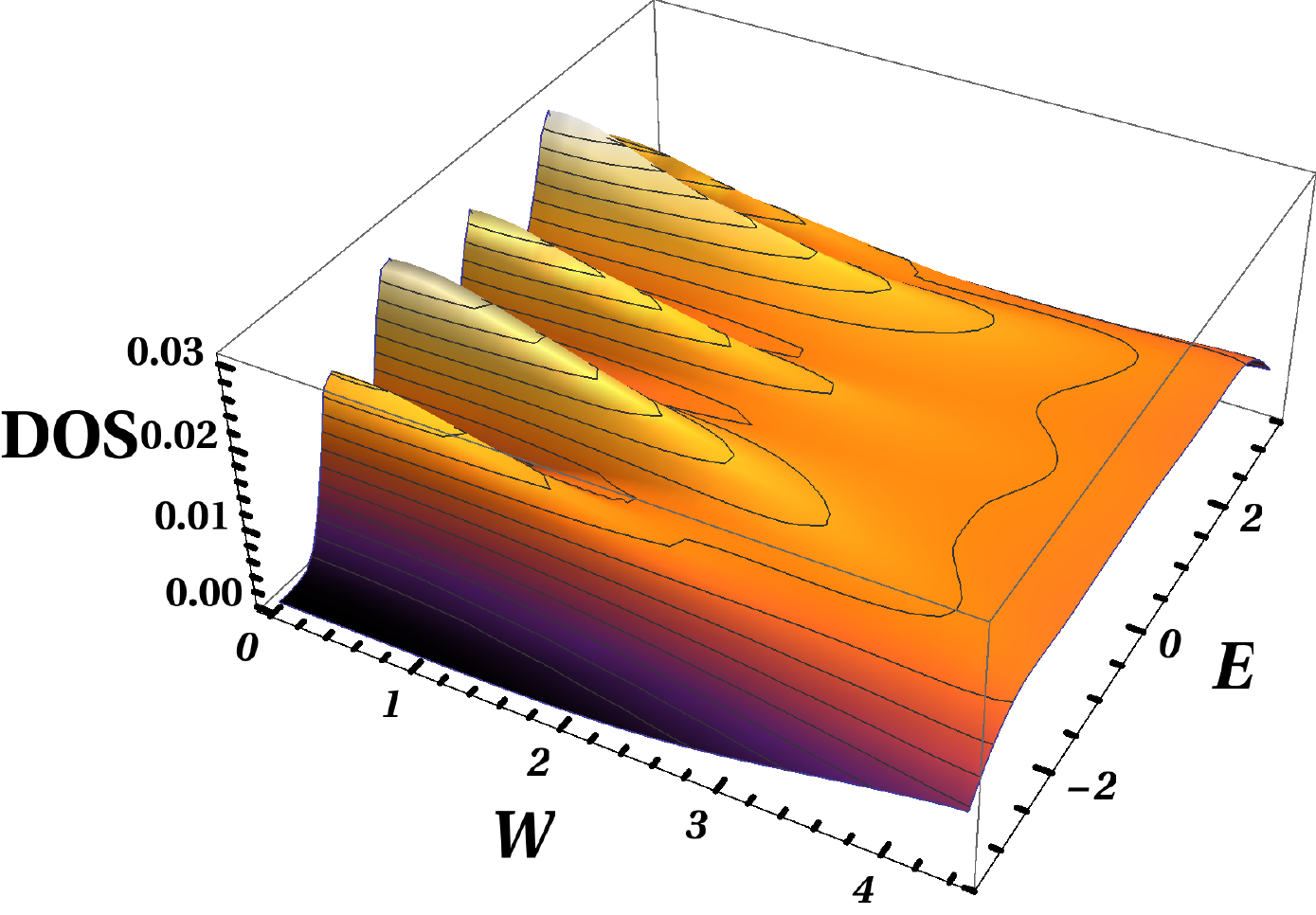}
    \includegraphics[width=0.49\columnwidth]{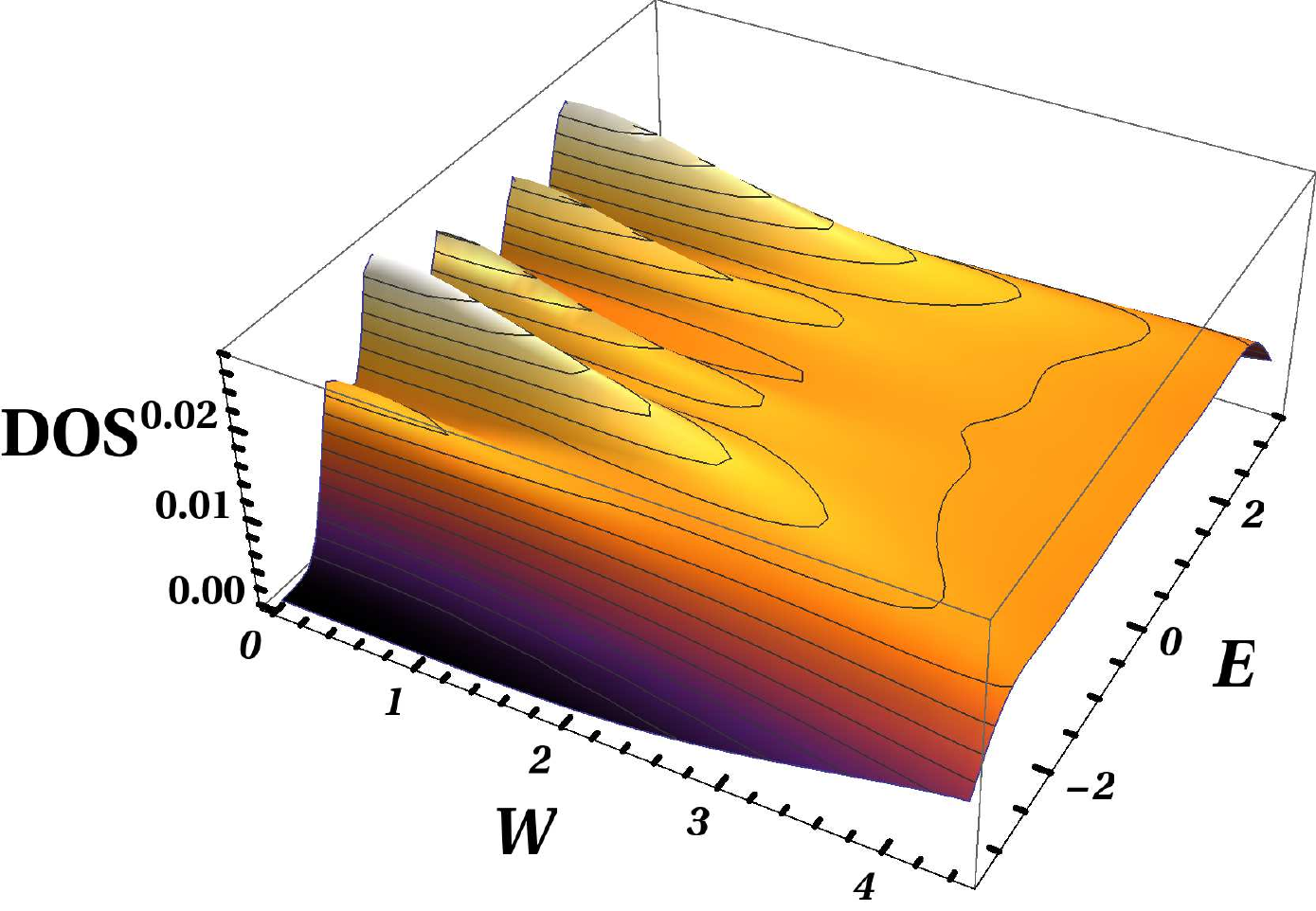}
    \caption{The band structure (top row) and the smoothed and normalized density of states (DOS, bottom row) for $\mathcal{L}(1)$ with width $M = 13$, $\mathcal{L}(2)$ with $M = 10$, $\mathcal{L}(3)$ with $M = 9$ and $\mathcal{L}(4)$ with $M = 8$. Colours in the top row vary from purple at low energies to red at high energies. In the bottom row, the colour indicates the numerical values of the DOS and the thin black lines are equal-DOS contours.}
    \label{fig:21_22_23_24_BandStructure_DOS}
\end{figure*}

\section{\label{sec:results}Results}

\subsection{\label{sec:l2x-dispersiondos}Dispersion relations and disorder-broadened density of states for $\mathcal{L}(n)$}

In the clean lattice $\mathcal{L}(1)$, we can easily compute the dispersion relations from the tight-binding model \eqref{eqn:Hamiltonian} as
\begin{equation}
    E_1    = 0, \quad
    E_{2,3}=\pm \sqrt{4+2\left(\cos k_x +\cos k_y \right)}
\end{equation}
where we have chosen the unit length to be $1$, and the $k_x, k_y$ are the wave vectors mapping onto the $x$ and $y$ axes. Fig.\ \ref{fig:21_22_23_24_BandStructure_DOS} plots the $\mathcal{L}(1)$ band structure and we can clearly see the flat band at energy $E=0$, and a Dirac cone around the `$M$' point $(k_x,k_y)=(\pi, \pi)$.
Similarly, we compute the band structures for $\mathcal{L}(n)$, $n= 2, 3,4$ and show them in the last three plots in Fig.\ \ref{fig:LiebLatticeStructure}. We see that every $\mathcal{L}(n)$ exhibits $n$ flat bands separated by $n+1$ dispersive bands \cite{Zhang2017NewBands}. Furthermore, the odd $n=1,3$ lattices show Dirac cone at special points in the $(k_x,k_y)$ plane while the even $n=2,4$ lattices only seem to have parabolic-type dispersions.



We now turn on the disorder, i.e.\ $W> 0$. Obviously, $k_x$ and $k_y$ are no longer good quantum numbers and the dispersion relations lose their meaning. Nevertheless, as discussed previously, the flat bands correspond to zero kinetic energy are dominated by the remaining term in the Hamiltonian, i.e.\ the random disorder term. Hence already small $W$ values can quickly lift the degeneracy of the flat bands and the states can begin the overlap with states in the originally dispersive, neighboring bands. 
In order to study the interplay of flat bands and dispersive bands in the disordered situation, we have computed the disorder-dependence of the density of states (DOS) by exact diagonalization for small system sizes $M^2= 13^2, 10^2, 9^2$ and $8^2$ for $\mathcal{L}(1)$, $\mathcal{L}(2)$, $\mathcal{L}(3)$ and $\mathcal{L}(4)$, respectively. The DOS has been obtained after averaging over $300$ 
samples. The results are given in Fig.\ \ref{fig:21_22_23_24_BandStructure_DOS}.
Clearly, the figure shows that in all cases the degeneracy of the flat bands is lifted very quickly and the originally degenerate states rapidly begin to move into energy regions occupied by the dispersive bands. Similarly, but somewhat less pronounced, the flat regions of the dispersive bands vanish upon increasing $W$. Overall, when $W$ becomes of the order of $2$--$3$, comparable to the band width of the largest subband in each case, the DOS loses it peak features.



\subsection{\label{sec:localization}Localization and finite-size scaling}

\begin{figure*}[tb]
    \centering
    $\mathcal{L}(1)$\includegraphics[width=0.95\columnwidth]{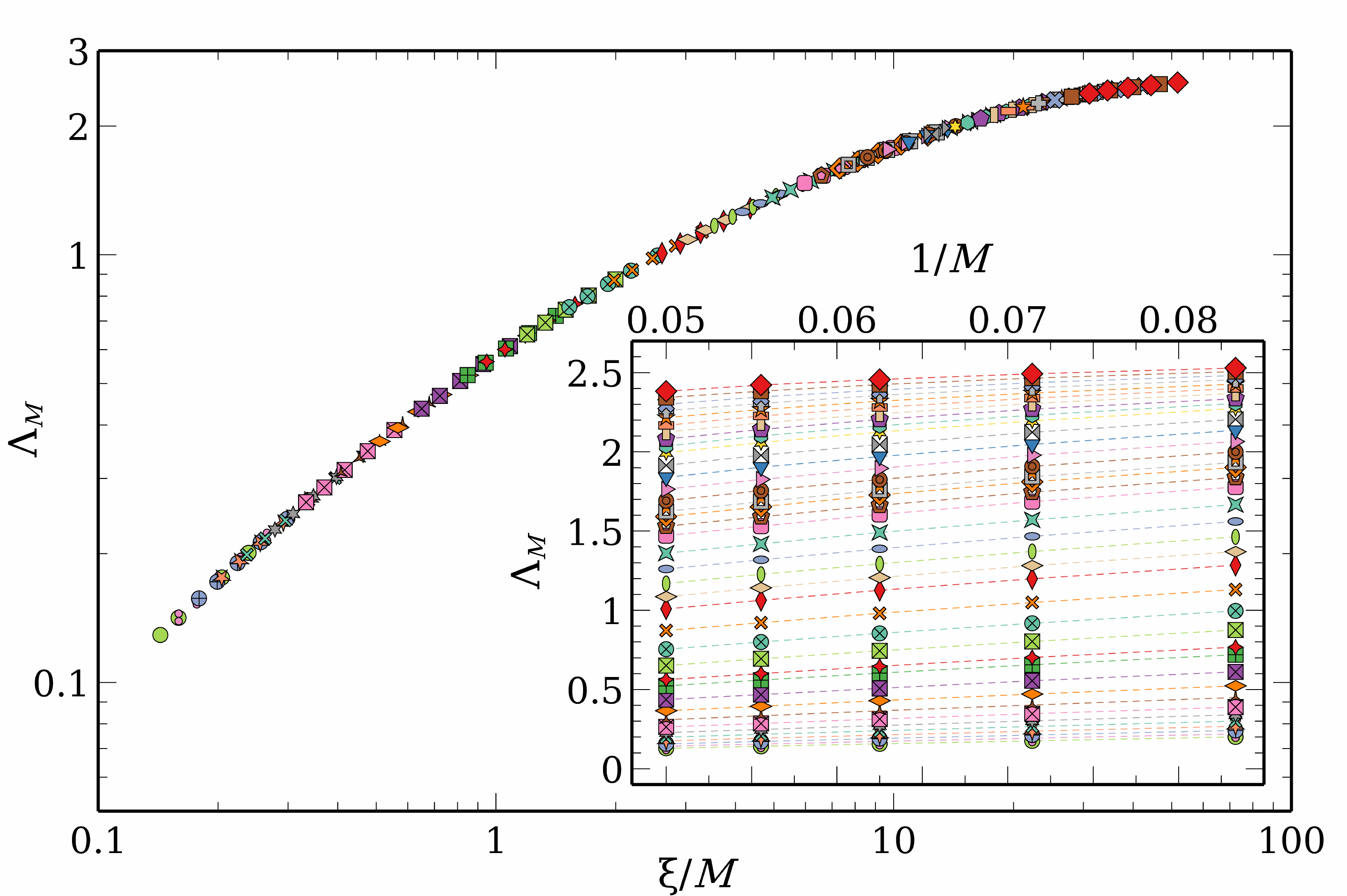}
    $\mathcal{L}(2)$\includegraphics[width=0.95\columnwidth]{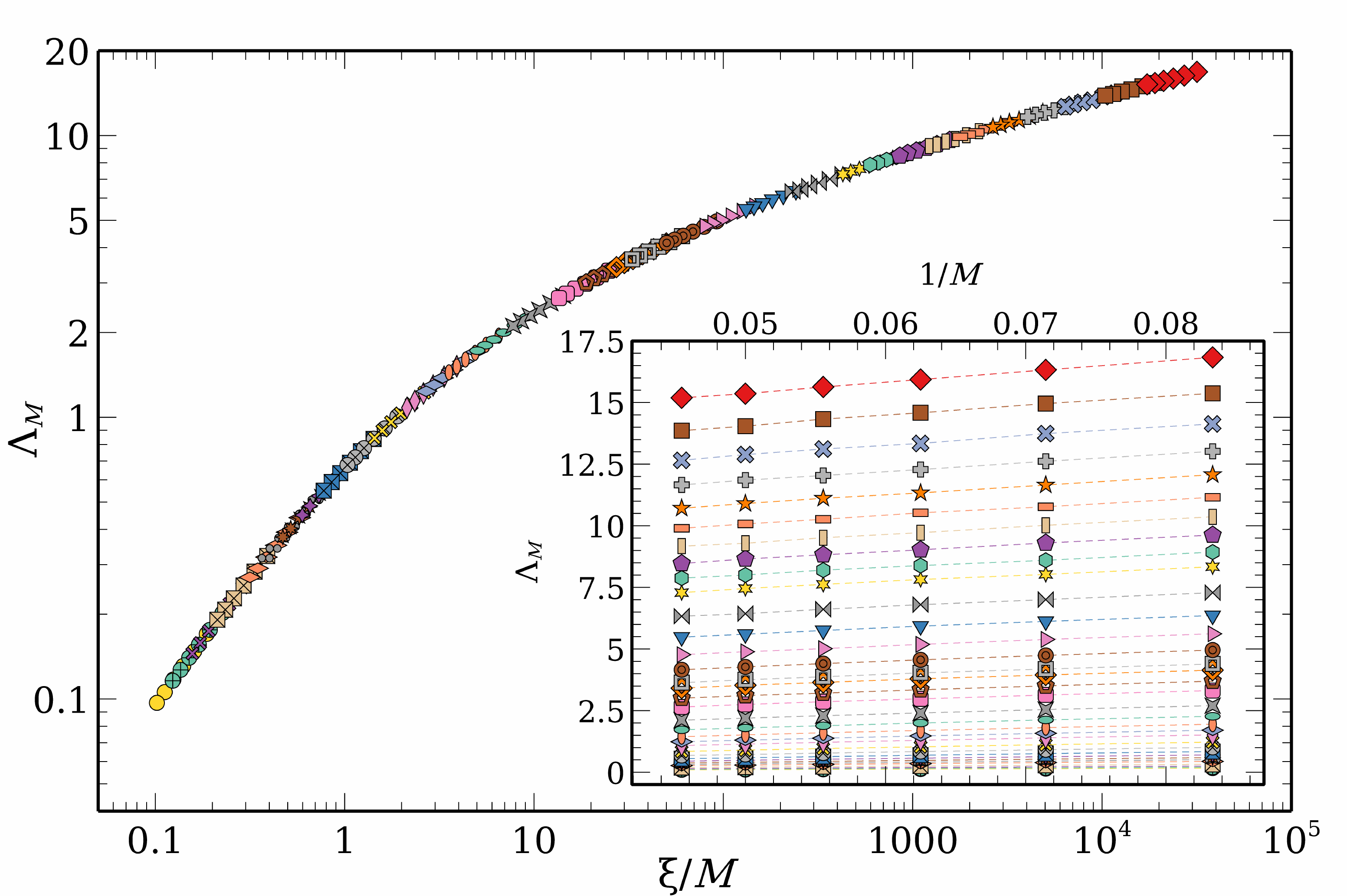}\\
    $\mathcal{L}(3)$\includegraphics[width=0.95\columnwidth]{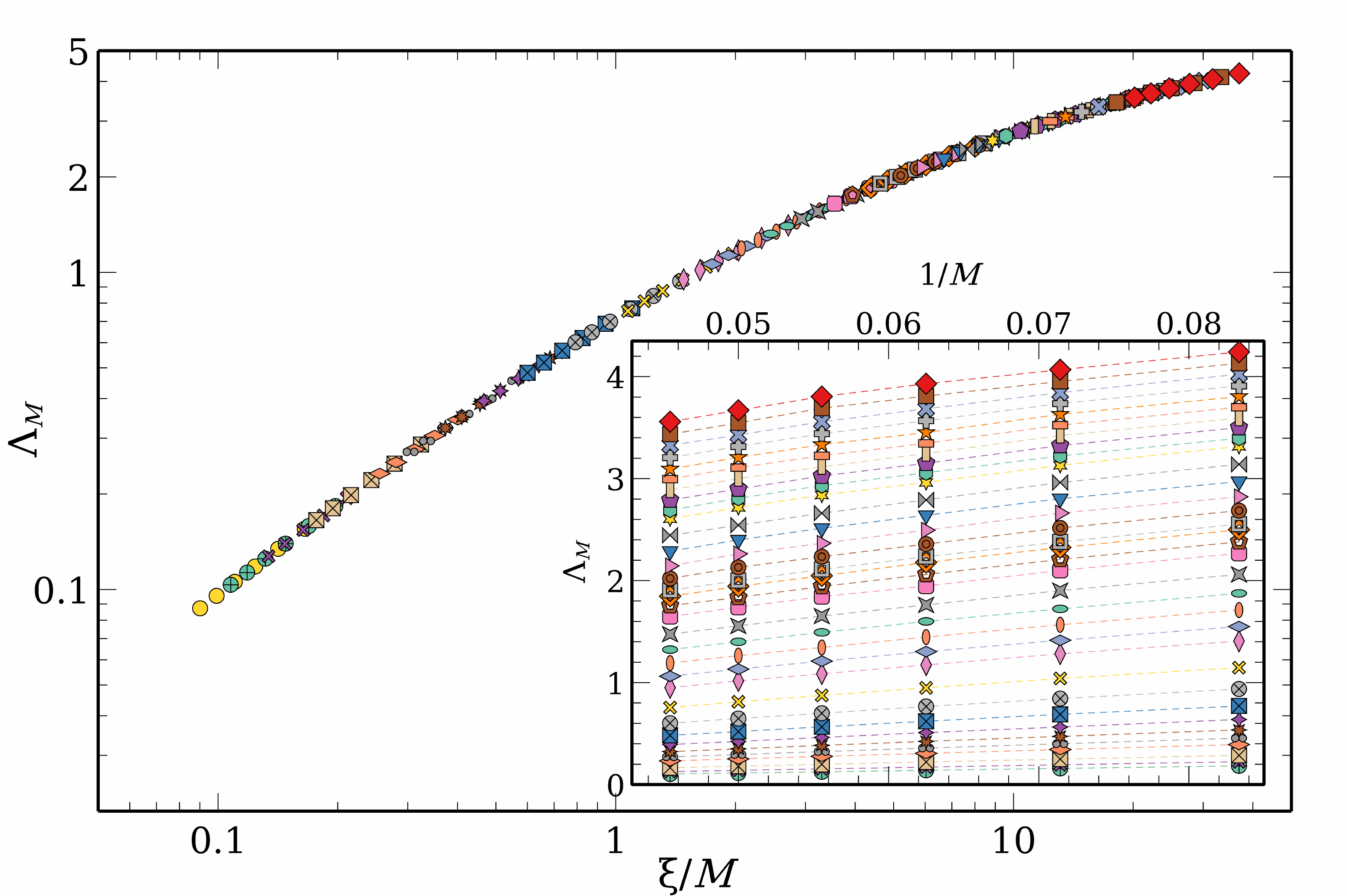}
    $\mathcal{L}(4)$\includegraphics[width=0.95\columnwidth]{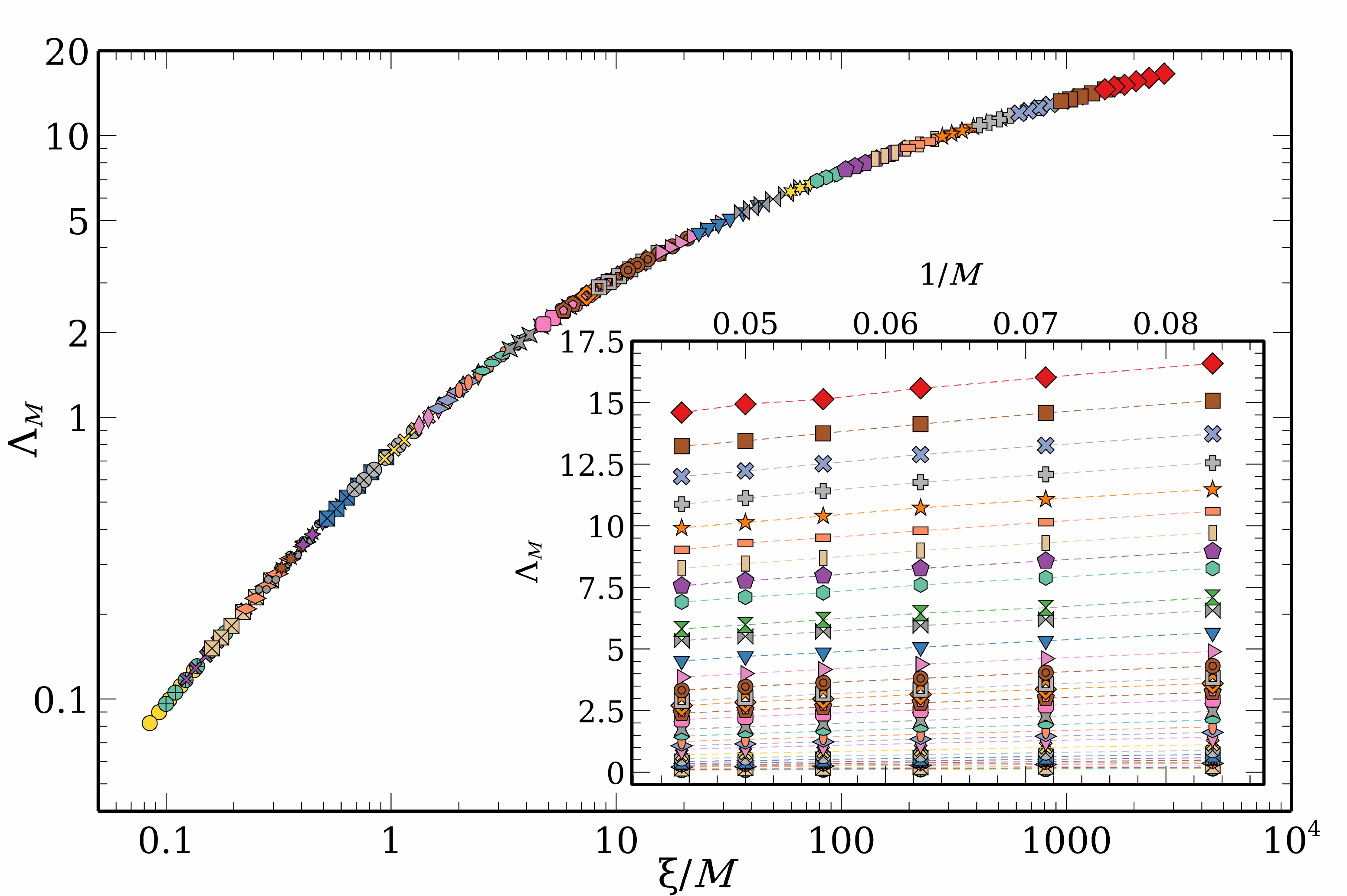}
    \caption{Double logarithmic plot of scaled reduced localization length, $\Lambda_M$, as a function of $\xi/M$ for $\mathcal{L}(1)$, $\mathcal{L}(2)$, $\mathcal{L}(3)$ and $\mathcal{L}(4)$ as indicated. The chosen values for $W$ are $1.0$ ($\Diamond$), $1.01$, \ldots, $1.05$ ($\Box$), \dots, $2.0$, $2.1$, $2.2$, $2.4$, $2.6$ \ldots, $10.0$ ($\bigcirc$) and vary in color. Some data points are not shown with symbols for clarity. Error bars are within symbol size. Insets: $\Lambda_{M}$ as a function of $1/M$ for each $\mathcal{L}(n)$, respectively. Symbols and colors are as in the main panels, dashed lines are guides to the eye only.}
    \label{fig:Lieb2DScalingDiagrams}
\end{figure*}

In Fig.\ \ref{fig:Lieb2DScalingDiagrams}, we give the result of the TMM calculations for $\Lambda_M(E,W)$ at $E=0$. The \revisionB{transverse system widths} were $M=10, 12, \ldots, 20$ with error $0.1\%$ for $\mathcal{L}(1)$, and $M=10,12, \ldots, 22$ with error $0.2\%$ for $\mathcal{L}(2)$, $\mathcal{L}(3)$ and  $\mathcal{L}(4)$. This results in about $10^8$ transfer matrix multiplications for each pair $(E,W)$.
We have chosen $E=0$ since this corresponds to the flat bands for $\mathcal{L}(1)$ and $\mathcal{L}(3)$ while it is in the centre of the subbands for $\mathcal{L}(2)$ and $\mathcal{L}(4)$. Hence $E=0$ allows us to study directly the influence of the disorder on originally flat and on originally dispersive bands. 
We find that in all cases, irrespective of the flatness of the original clean lattice bands, the values of $\Lambda_M(0,W)$ reduce for all values of $W$ when increasing $M$ as shown in the inset of Fig.\ \ref{fig:Lieb2DScalingDiagrams}. This is the behaviour as expected for localized states where for sufficiently large $W$ and $M$, the values of $\lambda_M$ saturate such that $\Lambda_M$ decreases when $M$ increases. 
Indeed, in the main panels of Fig.\ \ref{fig:Lieb2DScalingDiagrams}, we see that all $\Lambda_M(0,W)$ data can be made to scale onto single scaling curves. These curves also decrease with increasing $M$, hence showing strong localization. Furthermore, for $\lambda_M \ll M$ such that $\Lambda_M \ll 1$, we find that $\Lambda_M \propto 1/M$ as expected for strong localization with $\lambda_M \sim \lambda_\infty$ (where $\lambda_\infty$ denotes the value of the localization length in the thermodynamic limit).
We also note immediately that the localization lengths for the states originating from flat bands are nearly one order of magnitude smaller than the corresponding $\Lambda_M(0,W)$ values for the originally dispersive bands. Hence, in perfect agreement with our aforementioned expectations, we see that the absence of kinetic energy in the flat bands for the clean systems translates into strong localization for the disordered systems.



\begin{figure*}[tb]
\centering
    $\mathcal{L}(1)$\includegraphics[width=0.43\columnwidth]{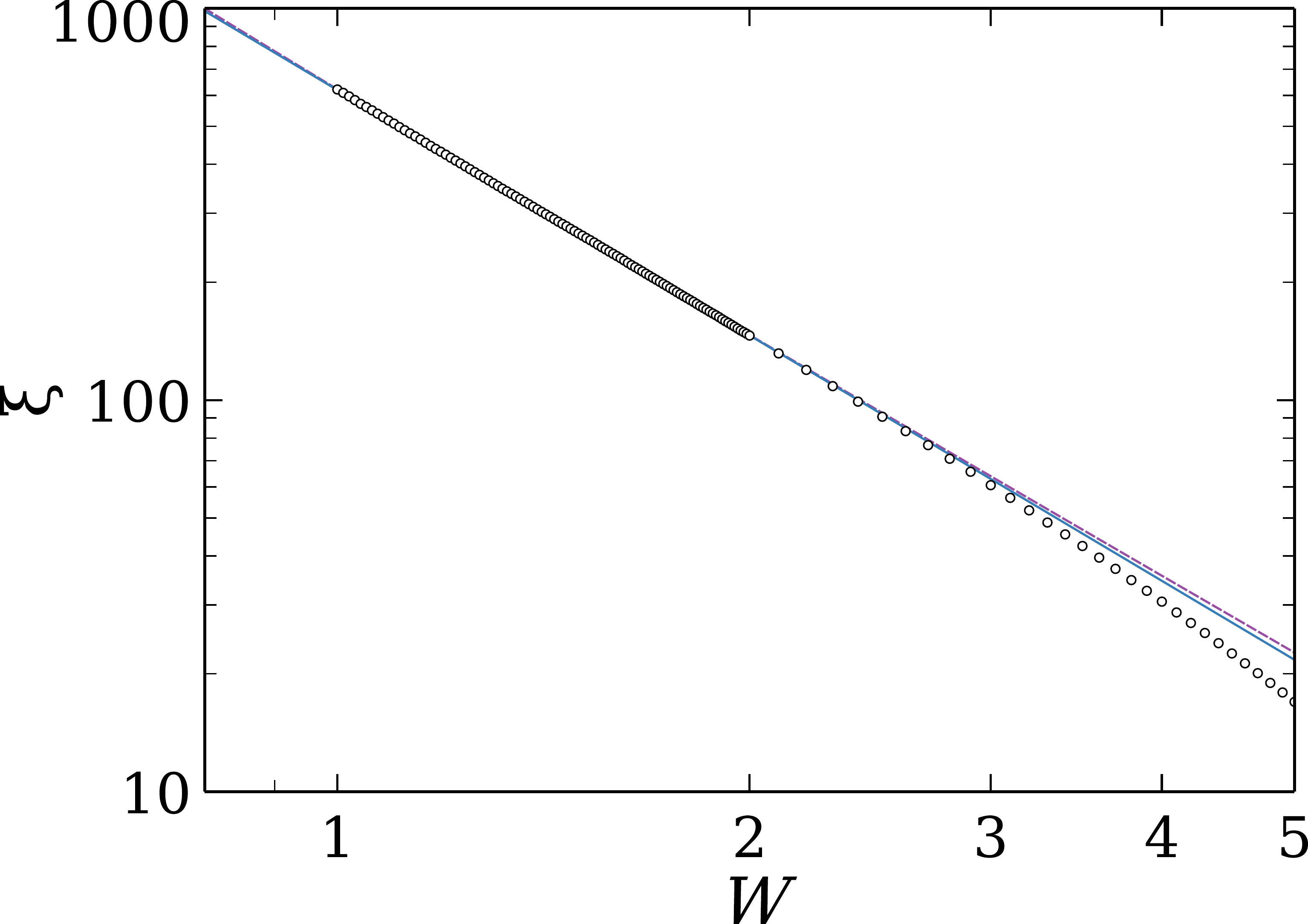}
    $\mathcal{L}(2)$\includegraphics[width=0.43\columnwidth]{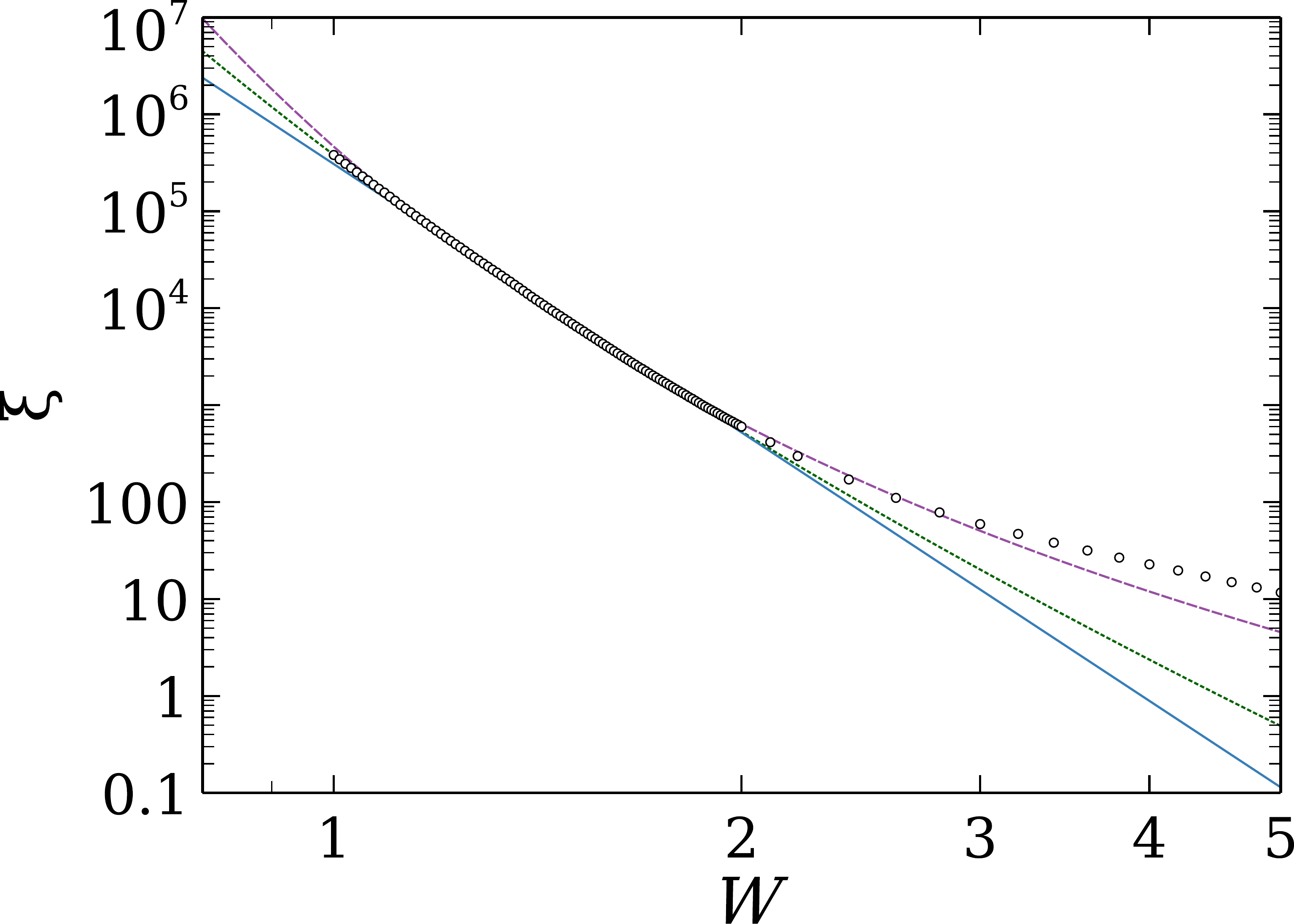}
    $\mathcal{L}(3)$\includegraphics[width=0.43\columnwidth]{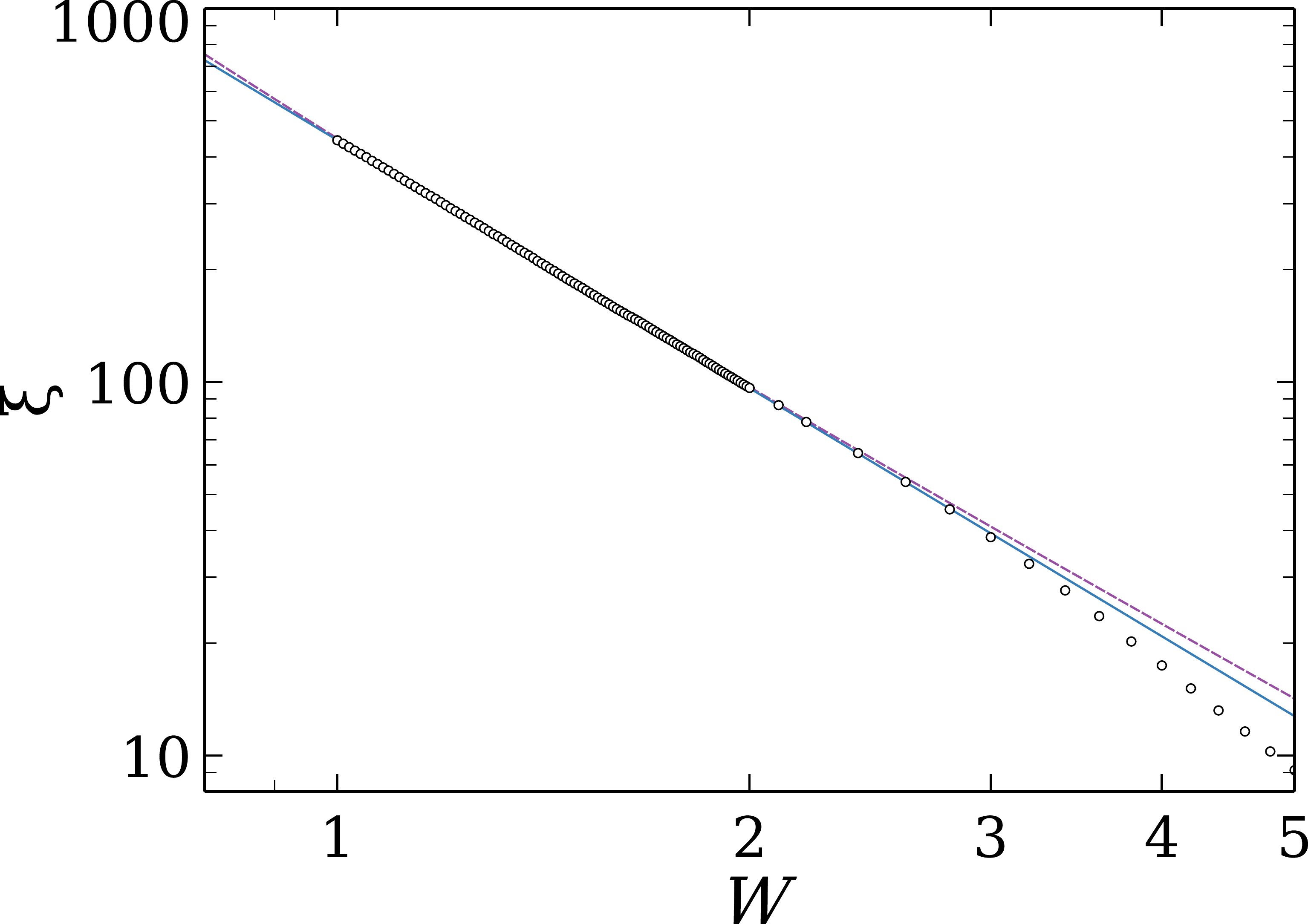}
    $\mathcal{L}(4)$\includegraphics[width=0.43\columnwidth]{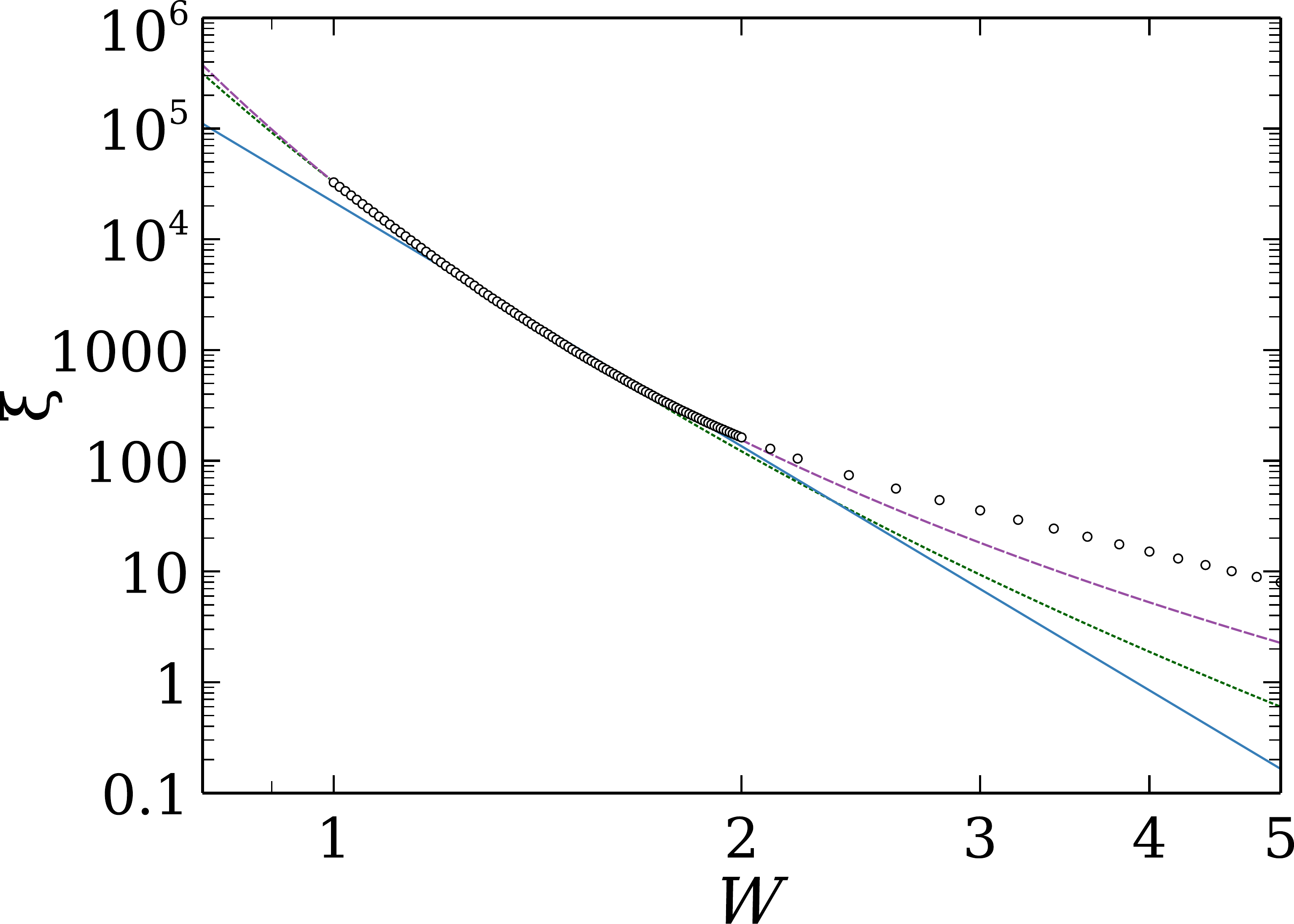}
     \caption{The scaling parameter $\xi(0,W)$ ($\bullet$) and fitted functional forms (lines) for $\mathcal{L}(1)$, $\mathcal{L}(2)$, $\mathcal{L}(3)$ and $\mathcal{L}(4)$, respectively. The blue solid line corresponds to $a W^{-\alpha}$, the purple dashed line is the fit function $a W^{-2} e^{\beta W^{-1}}$, while the green dotted line is the fit function $ a W^{-\alpha} e^{\beta {W^{-\gamma}}}$. The values for $a$, $\alpha$, $\beta$ and $\gamma$ are chosen from Tab.\ \ref{table:xiFitFunctionsFor2D}.}
    \label{fig:FitFunction}
\end{figure*}

In Fig.\ \ref{fig:FitFunction} we show the resulting $W$ dependence of the scaling lengths $\xi(0,W)$.
According to Thouless \cite{Edwards1972NumericalSystems,MacKinnon1983a}, we expect $\xi(W) \propto W^{-2}$ for a strictly 1D weakly disordered chain while in the 2D Anderson model the low $W$ behaviour has been shown to follow a non-universal $\xi(W)=a W^{-\alpha}\exp(\beta W^{-\gamma})$ with $a$, $\beta$ positive and $\alpha$, $\gamma$ of order unity \cite{Krameri1993}. These fits are also given in Fig.\ \ref{fig:FitFunction} and the resulting fit parameters are listed in Table \ref{table:xiFitFunctionsFor2D}. 
As is clear from Fig.\ \ref{fig:FitFunction}, the behaviour for $\mathcal{L}(1)$ and $\mathcal{L}(3)$, i.e.\ the flat bands, is clearly different from the behaviour of the dispersive bands for $\mathcal{L}(2)$ and $\mathcal{L}(4)$. This is already visible in the absolute magnitude of the $\xi$ values where the $\mathcal{L}(1)$ and $\mathcal{L}(3)$ cases are about three order of magnitude below the values for $\mathcal{L}(2)$ and $\mathcal{L}(4)$ for low $W$.
For $\mathcal{L}(1)$ and $\mathcal{L}(3)$, the simple power-law fit $\xi(W) \propto W^{-2}$ seems to work reasonably well at $1\leq W\leq 2$ while at $W>2$, in both cases the estimated $\xi$ values drop below the fit. One might want to speculate that this apparent consistency with a 1D fit result might also indicate that the spatial localization itself is along 1D structures, but this is clearly beyond what TMM can reliably infer. We also note that the agreement with the $\alpha=2$ value in 1D localization is not perfect while the value of $a$ should be $105$ at $E=0$ \cite{Kappus1981AnomalyModel} for $1D$ and about $12M$ for quasi-1D stripes of width $M$ \cite{ROMER2004}. Clearly, none of these values are obvious from the fit parameters given in Table \ref{table:xiFitFunctionsFor2D}.
On the other hand, for $\mathcal{L}(2)$ and $\mathcal{L}(4)$, none of our fits as given in Table \ref{table:xiFitFunctionsFor2D} seem very convincing with the non-universal 2D behavior perhaps performing slightly better. 
Of course the lack of a simple fit to the low-$W$ behaviour of $\xi$ is not surprising and simply indicates that the disordered Lieb models $\mathcal{L}(n)$ have indeed localization behaviour different from the standard 2D Anderson model.
\begin{table}
\centering
\begin{tabular}{llll}
\hline \hline
fit function &    parameter &     estimates &       $p$-value \\
\hline \hline 
\multicolumn{4}{c}{$\mathcal{L}(1)$}\\ \hline
        $a W^{-\alpha}$&                           $a$&           619.1(2)&       $<10^{-10}$\\   
        &                                       $\alpha$&       2.0816(7)&     $<10^{-10}$\\  
        $a W^{-2} e^{\beta W^{-1}}$&            $a$&        553(1)&         $<10^{-10}$\\  
        &                                       $\beta$&        0.117(2)&       $<10^{-10}$\\[1ex]         
                                
\multicolumn{4}{c}{$\mathcal{L}(2)$}\\ \hline
        $a W^{-\alpha}$&                           $a$&          310000(6000)&    $<10^{-10}$\\  
        &                                       $\alpha$&       9.20(4)&    $<10^{-10}$\\  
        $a W^{-\alpha} e^{\beta {W^{-\gamma}}}$&      $a$&        118(2000)&      0.94 \\ 
        &                                       $\alpha$&       5(5)&      0.26 \\
        &                                       $\beta$&        8(14)&      0.56 \\
        &                                       $\gamma$&       0.7(7)&    0.27 \\  
        $a W^{-2} e^{\beta W^{-1}}$&            $a$&            14.2(5)&    $<10^{-10}$\\
        &                                       $\beta$&        10.39(5)&    $<10^{-10}$\\[1ex]
        
\multicolumn{4}{c}{$\mathcal{L}(3)$}\\ \hline
        $a W^{-\alpha}$&                           $a$&            444.1(3)&       $<10^{-10}$\\   
        &                                       $\alpha$&       2.206(2)&      $<10^{-10}$\\  
        $a W^{-2} e^{\beta W^{-1}}$&            $a$&            334.9(6)&       $<10^{-10}$\\ 
        &                                       $\beta$&        0.292(3)&       $<10^{-10}$\\[1ex]
        
\multicolumn{4}{c}{$\mathcal{L}(4)$}\\ \hline
        $a W^{-\alpha}$&                           $a$&            21600(800)&     $<10^{-10}$\\   
        &                                       $\alpha$&       7.33(7)&       $<10^{-10}$\\  
        $a W^\alpha e^{\beta {W^{-\gamma}}}$&      $a$&            9(101)&        0.92 \\
        &                                       $\alpha$&       3(5)&          0.42 \\
        &                                       $\beta$&        8(11)&          0.45 \\
        &                                       $\gamma$&       0.8(6)&        0.16 \\  
        $a W^{-2} e^{\beta W^{-1}}$&            $a$&            11.5(2)&       $<10^{-10}$\\
        &                                       $\beta$&        7.96(3)&        $<10^{-10}$\\  \hline\hline
\end{tabular}
\caption{Fit functions for $\xi(W)$ and lattices $\mathcal{L}(1)$, $\mathcal{L}(2)$, $\mathcal{L}(3)$ and $\mathcal{L}(4)$ with $a$, $\alpha$, $\beta$, $\gamma$ are fit coefficients estimated from a Levenberg-Marquardt non-linear fitting procedure.}
\label{table:xiFitFunctionsFor2D}
\end{table}
\section{\label{sec:conclusions}Conclusions}
We have studied the localization properties of the disordered 2D Lieb lattice and its extensions. After appropriately finite-size scaling, we find that all states are localized, \revision{down to disorder of $W\sim 1$. For even lower disorders, we do not have sufficient computational resources for reach the same accuracies as presented here, but, as for the standard 2D Anderson model \cite{ROMER2004}, we also do not find any indications that the qualitative behaviour would change.}
Our results indicate that the finite-size scaled localization lengths $\xi$ for the energies corresponding to flat bands in the clean system ($W=0$) show a behaviour reminiscent of perturbative result for 1D. This could potentially provide spatial information about the nature of these localized states. For energies corresponding to dispersive bands in the clean system there is a tendency towards much larger localization lengths as is expected from the 2D Anderson model of localization \cite{Leadbeater1999}. Hence, the clear differences observed between flat and dispersive bands in photonic Lieb lattices \cite{Zhang2017NewBands} are predicted to be largely robust against the presence of disorder.


\section*{\label{sec:acknow}Acknowledgments}
We wish to acknowledge the National Natural Science Foundation of China (Grant No.\ 11874316), the Program for Changjiang Scholars and Innovative Research Team in University (Grant No.\ IRT13093), and the Furong Scholar Program of Hunan Provincial Government (R.A.R.) for financial support. This work also received funding by the CY Initiative of Excellence (grant "Investissements d'Avenir" ANR-16-IDEX-0008) and developed during R.A.R.'s stay at the CY Advanced Studies, whose support is gratefully acknowledged. We thank Warwick's Scientific Computing Research Technology Platform for computing time and support. UK research data statement: Data accompanying this publication are available from the corresponding authors.

\appendix
\section{\label{sec:appendix}Expressions for the dispersion relations}

For completeness, we give the dispersion relations for the Lieb lattices $\mathcal{L}(2)$, $\mathcal{L}(3)$ and $\mathcal{L}(4)$ here.
For $\mathcal{L}(2)$,
we find
\begin{subequations}
\begin{eqnarray}
    E_{1,2}&= \pm 1,\quad
    &E_{3}  =  \rho_+ + \rho_-,\\
    E_{4}  &= \omega \rho_+ + \omega^2 \rho_-,\quad
    &E_{5}  =  \omega \rho_- + \omega^2 \rho_+,
\end{eqnarray}
\end{subequations}   
where the $\omega=\frac{-1+\sqrt{3}i}{2}$, $\rho_{\pm}= \sqrt[3]{q \pm \sqrt{q^2 - \left( \frac{5}{3}\right)^3}}$
and $q=\cos k_{x} + \cos k_{y}$.
For $\mathcal{L}(3)$, we have
\begin{subequations}
\begin{eqnarray}
    E_{1}  &= &0, \quad
    E_{2,3} =  \pm \sqrt{2},\\
    E_{4,5,6,7}&= &\pm \sqrt{3\pm \sqrt{5+2 \left( \cos k_{x} + \cos k_{y}  \right)}}
    .
\end{eqnarray}
\end{subequations}
For $\mathcal{L}(4)$, the flat band solutions are given by
\begin{subequations}
\begin{eqnarray}
    E_{1,2,3,4}   &= &\frac{1}{2}\left( \pm 1\pm \sqrt{5}\right)
    .
\end{eqnarray}
The remaining $5$ dispersive bands emerge as solutions to a 5th order equation, i.e.\ 
\begin{equation}
E^5 - 7E^3 + 9E -2\left( \cos k_{x}+ \cos k_{y}\right) = 0,
\end{equation}
which we solve numerically.
\end{subequations}

\begin{thebibliography}{50}

\bibitem{Ashcroft1976}
N.~W. Ashcroft and N.~D. Mermin, {\em {Solid State Physics}} (Saunders College,
  New York, 1976).

\bibitem{Hook1991}
J.~R. Hook and H.~E. Hall, {\em {Solid State Physics}} (John Wiley
  {\textbackslash}{\&} Sons Ltd., The Atrium, Southern Gate, Chichester, PO19
  8SQ, 1991).

\bibitem{Kane2005}
C.~L. Kane and E.~J. Mele, Physical Review Letters {\bf 95},    (2005).

\bibitem{Bernevig2006}
B.~A. Bernevig, T.~L. Hughes, and S.-C. Zhang, Science {\bf 314},  1757
  (2006).

\bibitem{Wan2011TopologicalIridates}
X. Wan, A.~M. Turner, A. Vishwanath, and S.~Y. Savrasov, Physical Review B -
  Condensed Matter and Materials Physics {\bf 83},    (2011).

\bibitem{Xu2015ObservationMetal}
S.~Y. Xu {\it et~al.}, Science {\bf 347},  294  (2015).

\bibitem{Soluyanov2015Type-IISemimetals}
A.~A. Soluyanov {\it et~al.}, Nature {\bf 527},  495  (2015).

\bibitem{Leykam2018Perspective:Flatbands}
D. Leykam and S. Flach, APL Photonics {\bf 3},  070901  (2018).

\bibitem{Tasaki1998FromModel}
H. Tasaki, Progress of Theoretical Physics {\bf 99},  489  (1998).

\bibitem{Miyahara2007BCSLattice}
S. Miyahara, S. Kusuta, and N. Furukawa, Physica C: Superconductivity {\bf
  460-462},  1145  (2007).

\bibitem{Bergman2008BandModels}
D.~L. Bergman, C. Wu, and L. Balents, Physical Review B - Condensed Matter and
  Materials Physics {\bf 78},    (2008).

\bibitem{Wu2007FlatLattice}
C. Wu, D. Bergman, L. Balents, and S. Das~Sarma, Physical Review Letters {\bf
  99},  070401  (2007).

\bibitem{Leykam2017LocalizationStates}
D. Leykam, J.~D. Bodyfelt, A.~S. Desyatnikov, and S. Flach, The European
  Physical Journal B {\bf 90},  1  (2017).

\bibitem{Shukla2018}
P. Shukla, Physical Review B {\bf 98},  184202  (2018).

\bibitem{Ramachandran2017}
A. Ramachandran, A. Andreanov, and S. Flach, Physical Review B {\bf 96},  1
  (2017).

\bibitem{Goda2006InverseFlatbands}
M. Goda, S. Nishino, and H. Matsuda, Physical Review Letters {\bf 96},
  (2006).

\bibitem{Qiu2016DesigningSurface}
W.-X. Qiu {\it et~al.}, Physical Review B {\bf 94},  241409  (2016).

\bibitem{Julku2016GeometricBand}
A. Julku {\it et~al.}, Physical Review Letters {\bf 117},  045303  (2016).

\bibitem{Chen2017Disorder-inducedLattices}
R. Chen, D.-H. Xu, and B. Zhou, Physical Review B {\bf 96},  205304  (2017).

\bibitem{Nita2013}
M. Niţ{\u{a}}, B. Ostahie, and A. Aldea,   (2013).

\bibitem{Sun2018ExcitationLattice}
M. Sun, I.~G. Savenko, S. Flach, and Y.~G. Rubo, Physical Review B {\bf 98},
  161204  (2018).

\bibitem{Bhattacharya2019FlatLattice}
A. Bhattacharya and B. Pal,   (2019).

\bibitem{Lieb1989TwoModel}
E.~H. Lieb, Physical Review Letters {\bf 62},  1201  (1989).

\bibitem{Mielke1993FerromagnetismModel}
A. Mielke and H. Tasaki, Communications in Mathematical Physics {\bf 158},  341
   (1993).

\bibitem{Vicencio2015a}
R.~A. Vicencio {\it et~al.}, Physical Review Letters {\bf 114},  245503
  (2015).

\bibitem{Mukherjee2015a}
S. Mukherjee {\it et~al.}, Physical Review Letters {\bf 114},  245504  (2015).

\bibitem{Guzman-Silva2014ExperimentalLattices}
D. Guzm{\'{a}}n-Silva {\it et~al.}, New Journal of Physics {\bf 16},    (2014).

\bibitem{Diebel2016ConicalLattices}
F. Diebel {\it et~al.}, Physical Review Letters {\bf 116},    (2016).

\bibitem{Baboux2016BosonicBand}
F. Baboux {\it et~al.}, Physical Review Letters {\bf 116},    (2016).

\bibitem{Taie2015CoherentLattice}
S. Taie {\it et~al.}, Science Advances {\bf 1},  e1500854  (2015).

\bibitem{Shen2010SingleLattices}
R. Shen, L.~B. Shao, B. Wang, and D.~Y. Xing, Physical Review B {\bf 81},
  041410  (2010).

\bibitem{Slot2017ExperimentalLattice}
M.~R. Slot {\it et~al.}, Nature Physics {\bf 13},  672  (2017).

\bibitem{Souza2009Flat-bandModel}
A.~M.~C. Souza and H.~J. Herrmann, Physical Review B {\bf 79},    (2009).

\bibitem{Chalker2010AndersonBands}
J.~T. Chalker, T.~S. Pickles, and P. Shukla, Physical Review B - Condensed
  Matter and Materials Physics {\bf 82},    (2010).

\bibitem{Nishino2007Flat-bandSystem}
S. Nishino, H. Matsuda, and M. Goda, Journal of the Physical Society of Japan
  {\bf 76},    (2007).

\bibitem{Flach2014DetanglingLattices}
S. Flach {\it et~al.}, EPL {\bf 105},    (2014).

\bibitem{Vidal2000InteractionPotential}
J. Vidal, B. Dou{\c{c}}ot, R. Mosseri, and P. Butaud, Physical Review Letters
  {\bf 85},  3906  (2000).

\bibitem{Vidal2001DisorderCages}
J. Vidal {\it et~al.}, Physical Review B - Condensed Matter and Materials
  Physics {\bf 64},  1553061  (2001).

\bibitem{Gulacsi2004ExactInteractions}
Z. Gul{\'{a}}csi, Physical Review B - Condensed Matter and Materials Physics
  {\bf 69},    (2004).

\bibitem{Gulacsi2010RoutePolymers}
Z. Gul{\'{a}}csi, A. Kampf, and D. Vollhardt, Physical Review Letters {\bf
  105},    (2010).

\bibitem{Zhang2017NewBands}
D. Zhang {\it et~al.}, Annals of Physics {\bf 382},  160  (2017).

\bibitem{MacKinnon1983a}
A. MacKinnon and B. Kramer, Zeitschrift f{\"{u}}r Physik B Condensed Matter
  {\bf 53},  1  (1983).

\bibitem{Krameri1993}
B. Kramer and A. MacKinnon, Reports on Progress in Physics {\bf 56},  1469
  (1993).

\bibitem{Milde2000a}
F. Milde, Ph.D. thesis, 2000.

\bibitem{Cheraghchi2005Localization-delocalizationDisorder}
H. Cheraghchi, S.~M. Fazeli, and K. Esfarjani, Physical Review B - Condensed
  Matter and Materials Physics {\bf 72},    (2005).

\bibitem{Slevin1999b}
K. Slevin and T. Ohtsuki, Physical Review Letters {\bf 82},  382  (1999).

\bibitem{psiAD} Strictly speaking, the symbols A and B, as well as C later, are not needed to label $\Psi^{A,B}_x$, but we retain them here for the readers convenience.

\bibitem{Oseledets1968ASystems}
V. Oseledets, Trans. Moscow Math. Soc. {\bf 19},  179  (1968).

\bibitem{Ishii1973LocalizationSystem}
K. Ishii, Progress of Theoretical Physics Supplement {\bf 53},  77  (1973).

\bibitem{Beenakker1997Random-matrixTransport}
C.~W. Beenakker, {Random-matrix theory of quantum transport}, 1997.

\bibitem{Wilson1974TheExpansion}
K.~G. Wilson and J. Kogut, {The renormalization group and the
  {$\epsilon$}{\{}lunate{\}} expansion}, 1974.

\bibitem{Wegner1976ElectronsEdge}
F.~J. Wegner, Zeitschrift f{\"{u}}r Physik B Condensed Matter and Quanta {\bf
  25},  327  (1976).

\bibitem{Lee1979Real-spaceLocalization}
P.~A. Lee, Physical Review Letters {\bf 42},  1492  (1979).

\bibitem{Sarker1981ScalingApproach}
S. Sarker and E. Domany, Physical Review B {\bf 23},  6018  (1981).

\bibitem{Leadbeater1999}
M. Leadbeater, R. R{\"{o}}mer, and M. Schreiber, The European Physical Journal
  B {\bf 8},  643  (1999).

\bibitem{Rodriguez2011MultifractalTransition}
A. Rodriguez, L.~J. Vasquez, K. Slevin, and R.~A. R{\"{o}}mer, Physical Review
  B {\bf 84},  134209  (2011).

\bibitem{Edwards1972NumericalSystems}
J.~T. Edwards and D.~J. Thouless, Journal of Physics C: Solid State Physics
  {\bf 5},  807  (1972).

\bibitem{Kappus1981AnomalyModel}
M. Kappus and F. Wegner, Z. Phys. B -Condensed Matter {\bf 45},  15  (1981).

\bibitem{ROMER2004}
R.~A. R{\"{o}}mer and H. Schulz-Baldes, Europhysics Letters (EPL) {\bf 68},
  247  (2004).


\end{thebibliography}

\end{document}